\documentclass[preprint, preprintnumbers, amsmath, amssymb, aps, nofootinbib,longbibliography]{revtex4-1}

\usepackage[T1]{fontenc}
\usepackage{natbib}
\usepackage[normalem]{ulem}

\usepackage{amssymb,amsmath,mathrsfs,amsbsy}
\usepackage{graphicx,multirow,xcolor,soul}
\setstcolor{red} \setul{}{1.5pt}
\sethlcolor{lime}

\newcommand{\ps}   [2]{\ensuremath{\left(\left. {#1} \, \right| \, {#2} \right) }}

\newcommand{\psw}  [2]{\ensuremath{\left\langle\left. {#1} \, \right| \, {#2} \right\rangle }}

\newcommand{\ddfrac}[2]{\ensuremath{\frac{ \mathrm{d}{#1} }{ \mathrm{d}{#2} }}}

\newcommand\Real{\mbox{Re}} 

\newcommand\Rey{\mbox{\textit{Re}}} 
\newcommand\be{\begin{equation}} 
\newcommand\ee{\end{equation}}
\newcommand{\BF}[1]{\mathbf{#1}}
\def\00{\mathbf{0}}
\def\AAA{\mathbf{A}}
\def\aa{\mathbf{a}}
\def\bb{\mathbf{b}}
\def\CC{\mathbf{C}}
\def\ddd{{\partial}}
\def\ex{\mathbf{e}_x}
\def\ff{\mathbf{f}}
\def\II{\mathbf{I}}

\def\nn{\mathbf{n}}
\def\nnu{\Rey^{-1}}
\def\PP{\mathbf{P}}

\def\QQ{\mathbf{Q}}
\def\qq{\mathbf{q}}

\def\Ua{U^\dag} 	
\def\UUa{\mathbf{U}^\dag}		
\def\UU{\mathbf{U}}		

\def\Va{V^\dag} 
\def\Ut{\widetilde{U}}
\def\Vt{\widetilde{V}}
\def\Wt{\widetilde{W}}

\providecommand\bdelta{\boldsymbol{\delta}}
\providecommand\bnabla{\boldsymbol{\nabla}}
\providecommand\bcdot{\boldsymbol{\cdot}}
\def\cbz{\cos(\beta z)}
\def\sbz{\sin(\beta z)}

\def\Oetwo{O\left(\epsilon^2\right)}
\def\Oethree{O\left(\epsilon^3\right)}

\begin{document}

\title{Optimal spanwise-periodic control for recirculation length in a backward-facing step flow}

\author{E. Yim}
\affiliation{LFMI, \'Ecole Polytechnique F\'ed\'erale de Lausanne, CH-1015 Lausanne, Switzerland}

\author{I. Shukla}
\affiliation{LFMI, \'Ecole Polytechnique F\'ed\'erale de Lausanne, CH-1015 Lausanne, Switzerland}

\author{F. Gallaire}
\affiliation{LFMI, \'Ecole Polytechnique F\'ed\'erale de Lausanne, CH-1015 Lausanne, Switzerland}

\author{E. Boujo}
\affiliation{LFMI, \'Ecole Polytechnique F\'ed\'erale de Lausanne, CH-1015 Lausanne, Switzerland}
\date{\today}

\begin{abstract}
Three-dimensional control is considered in the flow past a backward-facing step (BFS).
The BFS flow at Reynolds number $Re=500$ (defined with the step height and the maximum inlet velocity) is two-dimensional and linearly stable but increasingly receptive to disturbances, with a potential for amplification as the recirculation length increases. 
We compute optimal spanwise-periodic control (steady wall blowing/suction or wall deformation) for decreasing the recirculation length, based on a second-order sensitivity analysis. 
Results show that wall-normal velocity control is always more efficient than wall-tangential control. The most efficient spanwise wavelength for the optimal control depends on the location: $\beta=0.6$ on the upper wall and $\beta=1$ on the upstream part of the lower wall. 
The linear amplification of the optimal control resembles the maximum linear gain, which confirms the link between recirculation length and amplification potential in this flow. 
Sensitivity predictions for blowing/suction amplitudes up to $O(10^{-3})$ and wall deformation amplitudes up to $O(10^{-2})$ are in good agreement with three-dimensional direct numerical simulations. 
 For larger wall deformation amplitudes, the flow becomes unsteady.
This study illustrates how the concept of second-order sensitivity and the associated optimization method allow for a systematic exploration of the best candidates for spanwise-periodic control.
\end{abstract}

\maketitle

\vspace{3cm}

\section{Introduction}
The flow over a backward facing step (BFS) is a quintessential example of a noise amplifier flow. 
Any small perturbation initially applied either decays in time or is progressively convected downstream of the perturbation source, letting the flow eventually return to its base flow configuration.  In terms of global linear stability properties,  the BFS flow for an expansion ratio of 2 was found globally stable to two-dimensional (2D) perturbations regardless of the Reynolds number. 
In contrast,  three-dimensional (3D)  perturbations 
periodic in the spanwise direction first become statically unstable, {for $Re \geq 714$ \mbox{\cite{Lanzerstorfer12}} ($Re \geq 748$ with a short inlet channel \mbox{\citep{Barkley02}})}, where the Reynolds number
$Re = U_{in} h/ \nu$ is defined with the maximum incoming velocity $U_{in}$,
the step height $h$ and the kinematic viscosity $\nu$. 
Despite their asymptotic decay, 2D perturbations can undergo large amplification in space and time due to  non-normal effects \citep{Marquet10}, in accordance with the locally convectively unstable nature of the flow {\mbox{\citep{Blackburn08}}}\citep{Boujo15}.

From a practical point of view, the flow over a BFS is of  importance since it serves as a prototype of several non-parallel flows in complex geometries such as in airfoils, cavities  diffusers{, and combustors} {\mbox{\citep{McManus90,McManus91,Ghoniem02}}}. 
The BFS geometry facilitates the study of both the flow separation and the flow reattachment, thus incorporating the two most prominent features of separated flows. While several techniques based on a practical approach exist for flow control in such geometries, the application of the theory of optimal flow control to separated flows has only started quite recently.

Among the empirical flow control approaches, the use of spanwise-periodic structures is particularly promising. In the context of flow separation, \cite{pujals2011} have demonstrated that using arrays of suitably shaped cylindrical roughness elements, streaks can be artificially forced on the roof of a generic car model, the so-called Ahmed body, which suppress
the separation around the rear-end.  
More generally,  spanwise wavy modulations have been recognized, mainly through an iterative trial and error method, as an efficient method of control in several flow configurations: for flows past bluff bodies to regulate vortex shedding \citep{Tanner1972method, zdravkovich1981review,
tombazis1997study, bearman1998reduction, choi2008control}, for circular cylinders \citep{ahmed1992transverse, ahmed1993experimental, lee2007experimental,
lam2008large, zhang2016numerical}, for rectangular cylinders {\citep{lam2012numerical}} and in airfoils \citep{lin2013numerical, serson2017direct}, to name a few.

The effectiveness of steady spanwise waviness to control nominally two-dimensional flows has been rationalized through the generalization of linear sensitivity analysis \citep{hill1992theoretical, marquet2008sensitivity} to second order. 
In the case of spanwise-periodic control of 2D flows, the linear sensitivity indeed vanishes at first order and the leading-order variation eventually depends quadratically on the 3D control amplitude \citep{hinchperturbation, cossu2014stabilizing, Boujo15b}. 
This dependence has been already established through the works of \citet{Hwang13,del2014optimalB,del2014optimalA,del2014stabilizing} and \citet{tammisola2014second}. 
The control effectiveness relies on two main features: the linear amplification potential of spanwise-periodic disturbances through amplification mechanisms like the lift-up mechanism, and the quadratic sensitivity of the flow on the resulting flow modifications.

In this study, we use the reattachment length as proxy for the noise amplifying potential of the separated flow in conjunction with a quadratic sensitivity analysis.  
The significance of the reattachment location as an indicator of the flow stability has already been substantiated through the works of \citet{sinha1981laminar} and \citet{armaly1983experimental}. 
More recently, \citet{Boujo14b, Boujo15} investigated the link between recirculation length and stability properties in separated flows. 
They found that the reattachment point was highly sensitive to the control, with its sensitivity map deeply resembling that of the backflow area and recirculation area. 
Further, these three sensitivity maps resembled closely that of the optimal harmonic gain, implying that the flow becomes a weaker amplifier as the recirculation length decreases, i.e. as the reattachment point moves upstream. 
{The presence of an upper wall and the appearance of a secondary recirculation region on that upper wall for $Re \gtrsim 275$ \mbox{\citep{Barkley02,Blackburn08}} tend to increase the overall spatial amplification. In this paper, we focus on the primary recirculation region on the lower wall. 
}

In this direction, we aim to exploit the amplification potential of the stable flow in a 3D BFS to design optimal control strategies, such that the smallest required control amplitude is capable of influencing the recirculation strength, here quantified by the recirculation length. 
We thereby build on the framework of \citet{Boujo19}, designed to control optimally the growth rate of a nominally 2D flow using steady spanwise-periodic perturbations, which we extend here to the optimal quadratic control of the recirculation length. 
We derive a second-order sensitivity tensor, whose scalar product  with any small-amplitude control yields the modification in reattachment location.

\begin{figure}
\centerline{\includegraphics[width=\textwidth]{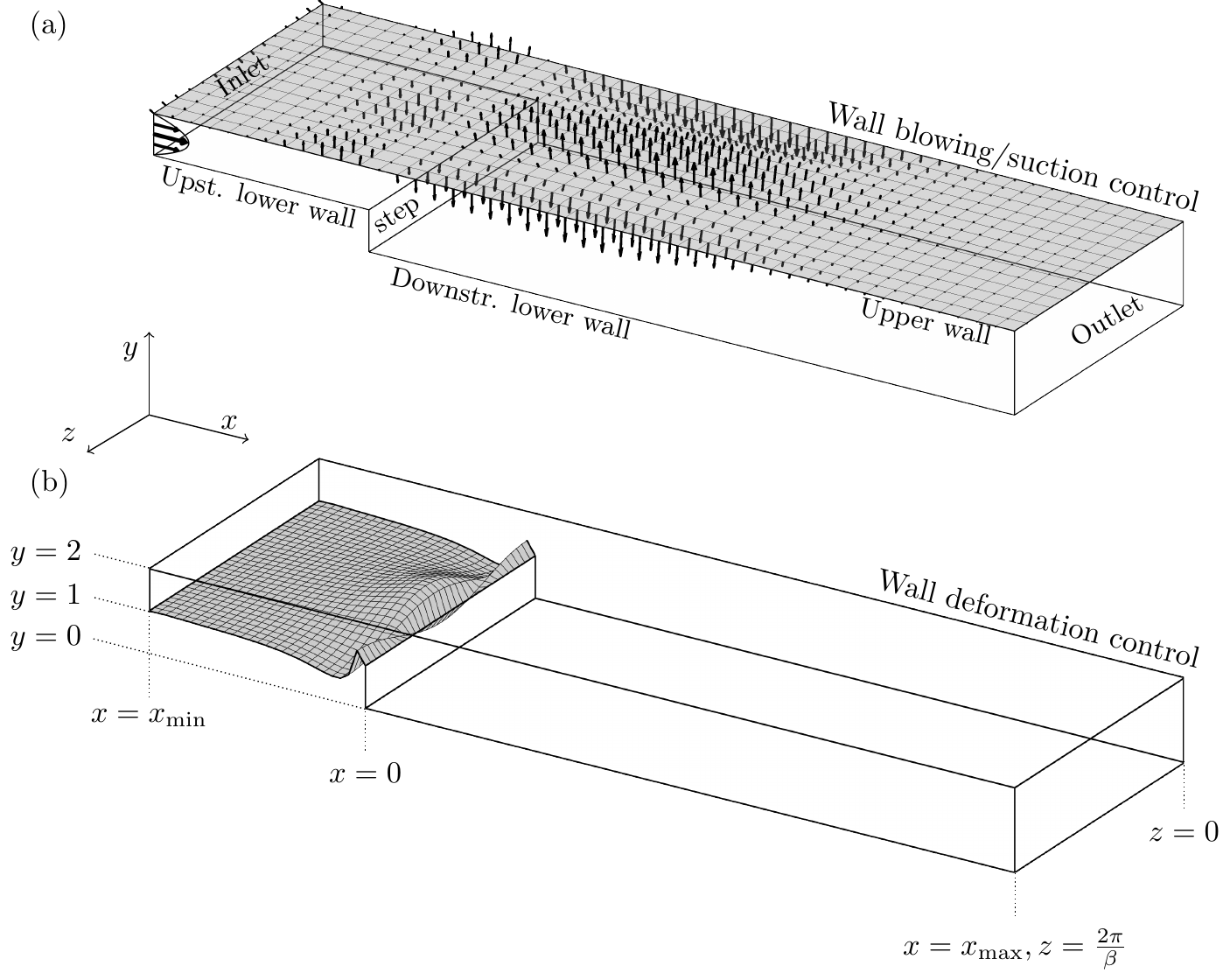}}
\caption{Sketches of steady spanwise-periodic control (wavenumber $\beta$) in a backward facing step: 
(a)~wall blowing/suction  applied on the upper wall and 
(b)~wall deformation  applied on the upstream lower wall.}
\label{fig:Sketch_control}
\end{figure}

Figure~\ref{fig:Sketch_control} shows the  optimal spanwise-harmonic control in a BFS  of expansion ratio 2. 
The geometry is bounded by $x\in[-5~50]$ and $y\in[0~2]$. 
The spanwise width is fixed at $z=[0~2\pi/\beta]$ where $\beta$ is the wavenumber of the control. 
We aim at optimizing the reattachment location 
using  wall actuation (Fig.~\ref{fig:Sketch_control}(a)) or wall deformation (Fig.~\ref{fig:Sketch_control}(b)). 
The Reynolds number is fixed at $Re=500$ throughout the analysis. 
This ensures that the flow is linearly stable to the steady 3D instability that occurs at  {$Re = 714$ ($Re = 748$ with a short inlet channel) with spanwise wavenumber $\beta=0.9$ \mbox{\citep{Barkley02,Lanzerstorfer12}}}. 

The paper is organized as follows. Section~\ref{SecPF} describes the problem formulation, the general expression of the second-order sensitivity tensor, and the optimization procedure used to compute the optimal control. 
Section~\ref{SecNM} presents
the numerical methods used for the sensitivity analysis and the optimization, as well as for  3D direct numerical simulations dedicated to validation. 
Global stability properties of the 2D uncontrolled flow are discussed  in  Sec.~\ref{SecGS}. 
The optimal wall actuation and wall deformation for minimizing the lower reattachment location are detailed in Sec.~\ref{SecRe}. 
We briefly discuss the limitations of the approach in Sec.~\ref{SecDiscussion}, before concluding in  Sec.~\ref{SecCo}.

\section{Problem formulation}\label{SecPF}
\subsection{Governing equations}
{Using $h$, $h/U_{in}$ and $\rho h^3$ as reference scales for length, time and mass, we}
 consider a steady 2D base flow $\BF{Q}(x,y) = (\UU, P)^T(x,y)=(U,V,P)^T(x,y)$ in a domain $\Omega$ of boundary $\Gamma$, that satisfies the {dimensionless} incompressible steady Navier-Stokes equations
\begin{align}
\bnabla  \cdot \UU = 0 \quad
\mathcal{N}(\BF{Q}) &= \mathbf{0} \quad \mathrm{in} \ \Omega, \\
\UU &= \mathbf{0}, \quad \mathrm{on} \ \Gamma, 
\end{align}
with $\mathcal{N}(\BF{Q}) \equiv \UU \cdot \bnabla \UU + \bnabla P -Re^{-1} \bnabla^2 \UU$, and $Re${$=U_{in} h/ \nu$} the Reynolds number {defined with the maximum incoming velocity $U_{in}$, the
step height $h$ and the kinematic viscosity $\nu$}. 

If there is a recirculation region, with  reattachment occurring on a wall defined by $y=y_w(x)$,
then the reattachment location $x_r$ is characterized by vanishing wall shear stress, 
\begin{equation} 
   \left. \frac{\partial U_t }{\partial n} \right|_{x=x_r, y=y_w(x_r)} = 0,
\end{equation}
i.e. vanishing normal derivative of the tangential velocity.
For the sake of simplicity, we now focus on the BFS flow: at the horizontal wall $y=0$, the reattachment location reduces to $\partial_y U (x_r,0) =0$;
in addition,  the flow separates at the step corner $x_s=0$, so the recirculation length  $l_c=x_r-x_s$ is simply $l_c=x_r$.

We assume that a 3D steady control of small amplitude $\epsilon$ is applied on a boundary $\Gamma_c$ with actuation velocity $\BF{U}_c(x,y,z)$, and possibly in the volume with body force  $\CC(x,y,z)$:
\begin{align}
 \bnabla \bcdot \UU = 0, \quad
\mathcal{N}(\BF{Q}) &= \epsilon \CC  \,\ \quad \mbox{ in } \Omega,
\\
 \UU&= \epsilon \UU_c \quad \mbox{ on } \Gamma_c,
  \\
 \UU&= \00  \,\qquad \mbox{ on } \Gamma\setminus\Gamma_c.
\end{align}
This 3D control modifies the 2D base flow as
\begin{alignat}{3}
\BF{Q}(x,y,z) &=\BF{Q}_0(x,y) &&+ \epsilon \BF{Q}_1(x,y,z) &&+ \epsilon^2 \BF{Q}_2(x,y,z) + \cdots,
\end{alignat}
where the $\QQ_i$ are solutions of the modified base flow equations at orders $\epsilon^0$, $\epsilon^1$ and $\epsilon^2$:
\begin{alignat}{3}
\mathcal{N}(\BF{Q}_0) & = \BF{0} && \quad \mathrm{in} \ \Omega, 
\quad \BF{U}_0= \BF{0} && \quad \mathrm{on} \  \Gamma,  \label{eq:OrderE0} 
\\
\BF{A}_0 \BF{Q}_1 & = (\BF{C},0)^T && \quad \mathrm{in} \ \Omega, 
\quad \BF{U}_1= \BF{U}_c &&  \quad \mathrm{on} \  \Gamma_c, 
\quad \BF{U}_1=\BF{0} \quad \mathrm{on} \ \Gamma \setminus \Gamma_c, \label{eq:OrderE1} 
\\
\BF{A}_0 \BF{Q}_2 & = (-\BF{U}_1 \cdot \nabla \BF{U}_1,0)^T && \quad \mathrm{in} \ \Omega, \quad \BF{U}_2= \BF{0} &&  \quad \mathrm{on} \ \Gamma, \label{eq:OrderE2}
\end{alignat}
and where $\AAA_0$ is the Navier-Stokes  operator linearized about the zeroth-order base flow $\QQ_0$,
 \begin{align}
{\AAA_0} &= \left[\begin{array}{cccc}
\UU_0 \bcdot\bnabla()    + () \bcdot \bnabla \UU_0    - \nnu \bnabla^2()  &   \bnabla ()  \\
 \bnabla \bcdot () & 0 
\end{array}\right].
\end{align}

The control and the resulting flow modification  alter the reattachment location as
\begin{equation}
x_r(z) = x_{r0} +\epsilon x_{r1}(z)  + \epsilon^2 x_{r2}(z) + \cdots.
\end{equation}
In this expression, $x_{r0}$ is the reattachment location of the uncontrolled flow $\QQ_0$,
\begin{equation}
   \left. \frac{\partial U_0 }{\partial y} \right|_{x=x_{r0}, y=0} = 0.
\end{equation}
Similarly, the first-order variation $x_{r1}(z)$ is the reattachment location of the first-order flow modification $\QQ_1$, characterized implicitly by a vanishing wall shear stress condition,
\begin{equation}
   \left. \frac{\partial U_1 }{\partial y} \right|_{x=x_{r1}, y=0,} = 0,
   \label{eq:xr1_impl}
\end{equation}
and expressed explicitly as  \citep{Boujo14, Boujo14b, Boujo15}:
\begin{equation}
   x_{r1}(z) = -\left.
   \frac{\partial_y U_1 }{\partial_{xy} U_0}
   \right|_{x=x_{r0}, y=0}.
   \label{eq:xr1_expl}
\end{equation}
The explicit dependence on $z$ in the notation $x_{r1}(z)$ in (\ref{eq:xr1_impl})-(\ref{eq:xr1_expl}) is meant to emphasize that the reattachment line is modulated in the spanwise direction.
When the control is harmonic in $z$, as considered in this study, it can actually be shown that $\QQ_1$ and $x_{r1}$ are purely harmonic too. 
As a result, the first-order variation $x_{r1}(z)$ has a zero mean.
In contrast, the second-order variation $x_{r2}(z)$ has a non-zero mean in general:
as detailed in Appendix~\ref{sec:app_xr2}, it reads
%
\begin{align}
   x_{r2}(z) &= \left[
   - \frac{\partial_y U_2 }{\partial_{xy} U_0} 
   +\frac{ \left(\partial_y U_1 \right) \left(\partial_{xy} U_1 \right)}{ \left( \partial_{xy} U_0 \right)^2}
   -\frac{ \left(\partial_{xxy} U_0\right) \left( \partial_y U_1 \right)^2 }{2 \left( \partial_{xy} U_0 \right)^3}
   \right]_{x=x_{r0}, y=0}
   \label{eq:xr2_1}
   \\
   &= x_{r2,\mathrm{I}} + x_{r2,\mathrm{II}} + x_{r2,\mathrm{III}}.
   \label{eq:xr2_2}
\end{align}
This expression shows that the reattachment location is modified at second order via two effects:
$x_{r2,\mathrm{I}}$ depends linearly on the second-order flow modification $\QQ_2$,
and 
$x_{r2,\mathrm{II}}$ and $x_{r2,\mathrm{III}}$ depend quadratically on the first-order flow modification $\QQ_1$.

\subsection{Sensitivity of the reattachment length: general expression}
\label{sc:sensit_general}

\begin{figure}
\centerline{\includegraphics[width=\textwidth]{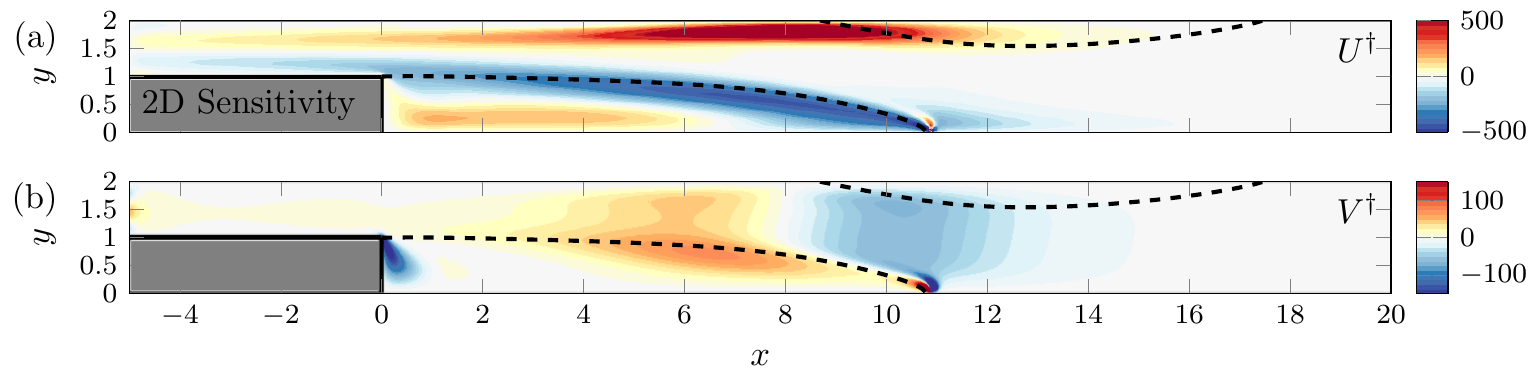}}
\caption{2D adjoint base flow (a) $U^\dagger$ and (b) $V^\dagger$. 
Dashed lines indicate lower and upper  recirculation regions, each of them delimited by a wall and a separating streamline (separatrix). }
\label{fig:Adjoint_base}
\end{figure}

We introduce the field ${\mathbf{S}}_\mathrm{I}$ and the operators 
${\mathbf{S}}_\mathrm{II}$  and
${\mathbf{S}}_\mathrm{III}$ 
such that
the second-order variation $x_{r2}$ 
can be expressed with scalar products,
\begin{align}
x_{r2}(z) = \ps{ {\mathbf{S}}_\mathrm{I} }{ \UU_2 }
+\ps{ \UU_1 }{ {\mathbf{S}}_\mathrm{II} \UU_1}
+\ps{ \UU_1 }{ {\mathbf{S}}_\mathrm{III} \UU_1},
\label{eq:1}
\end{align}
where the three terms of the right-hand side correspond to the three terms of (\ref{eq:xr2_1})-(\ref{eq:xr2_2}), respectively, and $\ps{\cdot}{\cdot}$ is the Hermitian scalar product in the domain $\Omega$ defined as 
$\ps{\aa}{\bb} \equiv \int_\Omega \BF{a}^*   \BF{b} \,\mathrm{d}\Omega$,
with the superscript $^*$ indicating complex conjugate. 
For integration along a boundary $\Gamma$, an angled bracket is used:  
$\psw{\aa}{\bb}\equiv \int_\Gamma \BF{a}^*  \BF{b} \,\mathrm{d}\Gamma$.
Omitting the notation $y=0$,
one identifies from (\ref{eq:xr2_1})-(\ref{eq:xr2_2}):
\begin{align}
   {\mathbf{S}}_\mathrm{I} &= \dfrac{ -1 }{\partial_{xy} U_0(x_{r0})} 
   \delta(x_{r0}) 
   \ex \partial_y,
   \label{eq:SI}
\\
   {\mathbf{S}}_\mathrm{II} &= 
   \dfrac{1}{ \left(\partial_{xy} U_0(x_{r0})\right)^2 }  
   \delta(x_{r0})
   \left( \ex \partial_y \right)^\dag
   \otimes
   \left( \ex \partial_{xy} \right),
   \label{eq:SII}
\\
   {\mathbf{S}}_\mathrm{III} &= 
   \frac{ - \partial_{xxy} U_0(x_{r0})  }{2 \left( \partial_{xy} U_0(x_{r0}) \right)^3}
   \delta(x_{r0})
   \left( \ex \partial_y \right)^\dag
   \otimes
   \left( \ex \partial_y \right),
   \label{eq:SIII}
\end{align}
where $\delta(x,y)$ is the 2D Dirac delta function,
and the superscript $^\dag$ denotes the adjoint of an operator defined as $\ps{\aa}{{\mathbf{S}}\bb} = \ps{{\mathbf{S}}^\dag\aa}{\bb}$.
Note that  ${\mathbf{S}}_\mathrm{I}$, ${\mathbf{S}}_\mathrm{II}$ and ${\mathbf{S}}_\mathrm{III}$ depend only on $\UU_0$.
From~(\ref{eq:OrderE2}), $\QQ_2$ is uniquely determined by $\QQ_1$, such that the first term of the right-hand side of (\ref{eq:1}) can be expressed as
\begin{align}
x_{r2,\mathrm{I}} &=
\ps{ {\mathbf{S}}_\mathrm{I} }{  -\AAA_0^{-1} (\UU_1 \bcdot  \bnabla \UU_1)}
=
\ps{{\AAA_0^\dag}^{-1} {\mathbf{S}}_\mathrm{I} }{ -\UU_1 \bcdot  \bnabla \UU_1}
= 
\ps{ \UUa }{ -\UU_1 \bcdot  \bnabla \UU_1}
\nonumber
\\
&=
\ps{\UU_1} {{\mathbf{S}}_\mathrm{I'} \UU_1},
\label{eq:2}
\end{align}
where we have introduced the 2D adjoint base flow $\UUa(x,y)$, defined by
\begin{align}
\AAA_0^\dag \UUa = {\mathbf{S}}_\mathrm{I},
\label{eq:Qa}
\end{align}
with $\AAA_0^\dag$  the adjoint Navier-Stokes operator.
The adjoint base flow, depicted in Fig.~\ref{fig:Adjoint_base},  depends only on $\UU_0$, and is the same adjoint base flow $\UUa$ as in \cite{Boujo14,Boujo15} where it represents the first-order sensitivity of the reattachment location $x_{r}$ to a steady 2D volume forcing.
In the last equality of (\ref{eq:2}), we were allowed to introduce an operator ${\mathbf{S}}_\mathrm{I'}$ (dependent on $\UUa$) because the expression is quadratic in $\UU_1$.
The second-order variation can therefore be expressed quadratically in any flow modification $\UU_1$ via a single operator for second-order sensitivity to flow modification:
\begin{align}
x_{r2}(z) &= \ps{\UU_1} {{\mathbf{S}}_{2,\UU_1} \UU_1}
\quad 
\mbox{where} 
\quad
{\mathbf{S}}_{2,\UU_1} = {\mathbf{S}}_\mathrm{I'} + {\mathbf{S}}_\mathrm{II} + {\mathbf{S}}_\mathrm{III}.
\end{align}
Finally, using (\ref{eq:OrderE1}), one can introduce operators for the second-order sensitivity  to control,  dependent only on the uncontrolled flow $\UU_0$, and such that for any control:

\begin{align}
x_{r2}(z) &= \ps{\CC} {{\mathbf{S}}_{2,\CC} \CC}
+ \psw{\UU_c} {{\mathbf{S}}_{2,\UU_c} \UU_c},
\end{align}
where 
\begin{align}
{\mathbf{S}}_{2,\CC} &= \PP^T  {\AAA_{0,\CC}^\dag}^{-1} {\mathbf{S}}_{2,\UU_1} {\AAA_{0,\CC}}^{-1} \PP,
\label{eq:S2C}
\\
\mbox{and }\quad
{\mathbf{S}}_{2,\UU_c} &= \PP^T  {\AAA_{0,\UU_c}^\dag}^{-1} {\mathbf{S}}_{2,\UU_1} {\AAA_{0,\UU_c}}^{-1} \PP.
\label{eq:S2Uc}
\end{align}
Here $\PP$ is the prolongation matrix that converts the velocity-only space to velocity-pressure space such that $\PP \UU = (\UU, 0)^T$ and  $\PP^T \QQ = \UU$, and  
$\AAA_{0,\CC}$ and $\AAA_{0,\UU_c}$ are defined by the volume-control-only and wall-control-only versions of~(\ref{eq:OrderE1}), respectively:
\begin{alignat}{3}
\AAA_{0,\CC} \BF{Q}_1 & = (\BF{C},0)^T && \quad \mathrm{in} \ \Omega, 
\quad \BF{U}_1= \00 &&  \quad \mathrm{on} \  \Gamma, 
\\
\AAA_{0,\UU_c} \BF{Q}_1 & = \00 && \quad \mathrm{in} \ \Omega, 
\quad \BF{U}_1= \BF{U}_c &&  \quad \mathrm{on} \  \Gamma_c, 
\quad \BF{U}_1=\BF{0} \quad \mathrm{on} \ \Gamma \setminus \Gamma_c.
\end{alignat}

\subsection{Simplification: spanwise-harmonic control}
Let us now assume a spanwise-harmonic control of the form 
\be 
\UU_c(x,y,z) =  \left( \begin{array}{c}
\widetilde U_{c}(x,y) \cbz \\ \widetilde V_{c}(x,y) \cbz \\ \widetilde W_{c}(x,y) \sbz
\end{array} \right),
\quad
\CC(x,y,z) =  \left( \begin{array}{c}
\widetilde C_x(x,y) \cbz \\ \widetilde C_y(x,y) \cbz \\ \widetilde C_z(x,y) \sbz 
\end{array} \right).
\label{eq:harm_U_C}
\ee
The first-order flow modification is also spanwise-harmonic, of same wavenumber $\beta$:
\begin{equation} \QQ_1(x,y,z) =
   \left(\begin{array}{c}
      \widetilde U_1(x,y) \cbz \\
      \widetilde V_1(x,y) \cbz \\
      \widetilde W_1(x,y) \sbz \\
      \widetilde P_1(x,y) \cbz
   \end{array}\right).
\end{equation}

The quadratic term $-\UU_1 \bcdot \bnabla\UU_1$ in (\ref{eq:OrderE2}) is then the sum of 
2D terms (spanwise-invariant terms, of wavenumber $0$) and 
3D terms (of wavenumber $2\beta$), which we denote $\ff^{2D}(x,y) + \ff^{3D}(x,y,z)$. 
As a result, the second-order flow modification has the same form: $\QQ_2^{2D}(x,y) + \QQ_2^{3D}(x,y,z)$.
Similarly, the second and third terms in (\ref{eq:xr2_1})-(\ref{eq:xr2_2}) and (\ref{eq:1}) have the same form too, and finally the second-order reattachment location modification reads 
\begin{equation}
   x_{r2}(z) = x_{r2}^{2D} + x_{r2}^{3D}(z)
\end{equation}
where
\begin{align}
   x_{r2}^{2D} &= \left[
   - \frac{\partial_y U_2^{2D} }{\partial_{xy} U_0} 
   +\frac{ \left(\partial_y \widetilde U_1 \right) \left(\partial_{xy} \widetilde U_1 \right)}{ 2\left( \partial_{xy} U_0 \right)^2}
   -\frac{ \left(\partial_{xxy} U_0\right) \left( \partial_y \widetilde U_1 \right)^2 }{4 \left( \partial_{xy} U_0 \right)^3}
   \right]_{x=x_{r0}, y=0}
   \label{eq:xr2_1_simpl}
   \\
   &= x_{r2,\mathrm{I}}^{2D} + x_{r2,\mathrm{II}}^{2D} + x_{r2,\mathrm{III}}^{2D}.
   \label{eq:xr2_2_simpl}
\end{align}
Because  $x_{r2}^{3D}(z)$ is harmonic of zero mean, we now focus on the spanwise-invariant component $x_{r2}^{2D}$. 
Its expression can be simplified,
taking advantage of the specific form (\ref{eq:harm_U_C}) of the control:

\begin{align}
x_{r2}^{2D} 
& =  \ps{\widetilde \CC} {\widetilde {\mathbf{S}}_{2,\widetilde \CC} \widetilde \CC}
+ 
\psw{\widetilde \UU_c} {\widetilde {\mathbf{S}}_{2,\widetilde \UU_c} \widetilde \UU_c},
\label{eq:S2_C_Uc}
\end{align}
where $\widetilde {\mathbf{S}}_{2,\widetilde \CC}$ and $\widetilde {\mathbf{S}}_{2,\widetilde \UU_c}$ are spanwise-invariant  versions of the second-order sensitivity operators (\ref{eq:S2C})-(\ref{eq:S2Uc})   (see detailed expressions   in Appendix~\ref{sec:app_Simplification}). 
The advantage of this simplification is that calculating the sensitivity operators (and, later, finding the optimal control) can be performed with 2D  fields and tensors, rather than 3D ones, which greatly reduces the computational cost and memory requirements.

\begin{figure}
\centerline{\includegraphics[width=\textwidth]{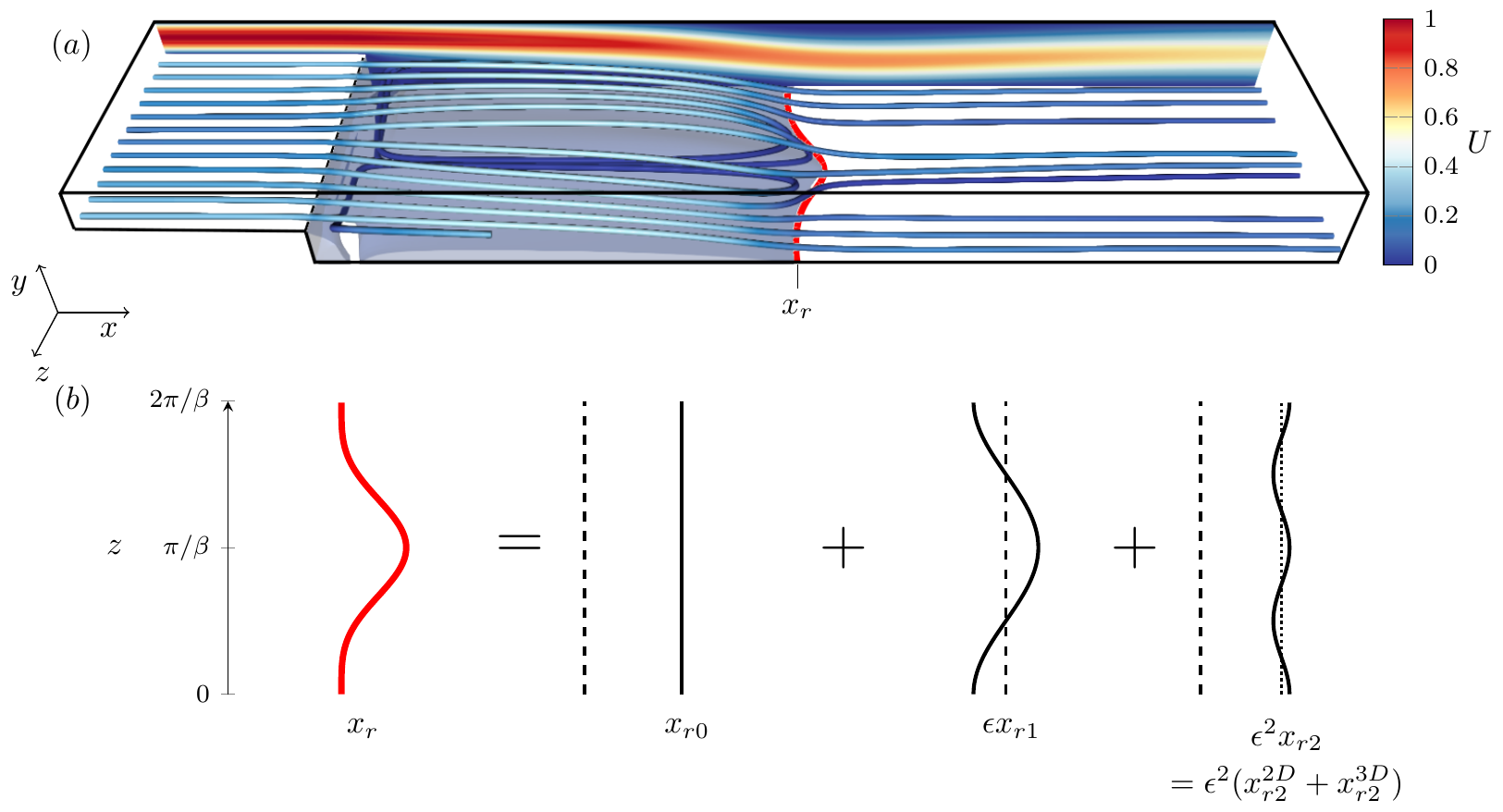}}
\caption{
(a) An example of 3D base flow modified by a wall blowing/suction control (using the same control as in Fig.~\ref{fig:Op_xr2d_v_b1} with $\epsilon =0.003$). 
Streamlines start at $(x,y)=(-5,1.05)$ at different spanwise positions $z$. 
The iso-surface indicates the lower zero streamwise velocity $U=0$  (the upper  recirculation region is not shown here). 
The thick red line indicates the lower reattachment location characterized by a vanishing wall shear stress $\partial_y U=0$. 
(b)~Decomposition of the reattachment location $x_r$ into zeroth, first and second-order components $x_{r0}$, $x_{r1}$ and $x_{r2}$. 
The spanwise-averaged reattachment location is $\overline{x_r} = x_{r0}+ \epsilon^2 x_{r2}^{2D}$.
}
\label{fig:Sketch_Xr}
\end{figure}

Figure~\ref{fig:Sketch_Xr}(a) visualizes a 3D flow obtained with spanwise-periodic control. 
The optimal wall normal blowing/suction control  for $\beta=1$ is applied  on the upstream part ($x<0$, $y=1$) of the lower wall, with amplitude $\epsilon=0.003$ (see  Fig.~\ref{fig:Op_xr2d_v_b1} for the actuation vector).  
As shown in the sketch of Fig.~\ref{fig:Sketch_Xr}(b),
the reattachment location $x_r(z)$ is decomposed into 
zeroth-order $x_{r0}$ (uncontrolled), 
first-order $x_{r1}(z)$ (of zero mean), and second-order $x_{r2}$. 
As mentioned earlier, the second-order component is further divided into a
zero-mean 3D part $x_{r2}^{3D}(z)$ and a mean 2D part $x_{r2}^{2D}$. 
Therefore, the spanwise-averaged  reattachment location is 
\begin{equation} 
   \overline{x_r} = x_{r0} + \epsilon^2 x_{r2}^{2D},
\end{equation} 
which is our control interest. 
The second-order variation $x_{r2}^{2D}$
is now referred to as mean correction.

\subsection{Optimal spanwise-periodic control}

In this section, we show how the spanwise-harmonic control can be optimized so as to yield the largest possible effect on the reattachment location. 
The formulation is similar to 
\cite{Boujo19}, where the control was optimized for the largest effect on the linear stability properties (growth rate or  frequency, i.e. real or imaginary part of the complex eigenvalue),
except that here all quantities are real. 
We only describe the optimization procedure for boundary control $\widetilde \UU_c$;  the derivation for volume control $\widetilde \CC$ is similar.

\subsubsection{Optimal spanwise-periodic wall actuation}

If the recirculation length is to be reduced,
the mean correction can be minimized by solving the following problem:
\begin{equation}
\min_{||\BF{\Ut_c}||=1} \left( x_{r2}^{2D} \right) 
=
\min \frac{\left< \BF{\Ut_c} \left| \frac{1}{2} \left(\widetilde{\mathbf{S}}_{2,\widetilde\UU_c}+\widetilde{\mathbf{S}}^{T}_{2,\widetilde\UU_c} \right) \BF{\Ut_c}  \right.  \right>}{\left< \BF{\Ut_c} \left| \BF{\Ut_c}  \right.  \right>} 
= 
\frac{1}{2} \lambda_{\min}  \left(\widetilde{\mathbf{S}}_{2,\widetilde\UU_c}+\widetilde{\mathbf{S}}^{T}_{2,\widetilde\UU_c} \right).
\label{eq:opUc}
\end{equation}
This indicates that, for any given wavenumber $\beta$, the smallest (largest negative) eigenvalue of the symmetric operator
 $\frac{1}{2} \left(\widetilde{\mathbf{S}}_{2,\widetilde\UU_c}+\widetilde{\mathbf{S}}^{T}_{2,\widetilde\UU_c} \right)$
 is the smallest (largest negative) mean correction, and the corresponding eigenvector $\UU_c$ is the optimal wall control.
 Similarly, if the recirculation length is to be increased, the mean correction can be maximized by finding the largest positive eigenvalue and the associated eigenvector.

\subsubsection{Optimal spanwise-periodic wall deformation}

For open-loop control, deforming the geometry can be more interesting than using a steady wall velocity actuation. 
It is possible to compute the optimal wall deformation, noting that  an equivalent wall deformation can be deduced from a given wall blowing/suction control \citep{Boujo19}. 
On wall boundaries, the velocity should vanish;
for a small-amplitude wall-normal deformation $\epsilon y_1$, this condition yields (with a Taylor expansion):
\begin{align}
\mathbf{U}(y_0+\epsilon y_1) &= \mathbf{U}_0(y_0+\epsilon y_1) + \epsilon\mathbf{U}_1(y_0+\epsilon y_1) + \cdots \nonumber \\
&=\mathbf{U}_0(y_0) + \epsilon \left[y_1 \partial_y \mathbf{U}_0(y_0)+ \mathbf{U}_1(y_0) \right]  +\cdots = \mathbf{0}.
\end{align}
Noting that $\UU_0(y_0)=\mathbf{0}$, this gives the relation between  wall-normal deformation $y_1$ and equivalent tangential velocity $U_c$:
\begin{equation}
U_1(y_0) = -y_1 \frac{\partial U_0(y_0)}{\partial y} = U_c.
\label{eq:walldef}
\end{equation}
Therefore, considering spanwise-harmonic wall-normal deformations of the form 
\be 
y_1(z) = \tilde y_1 \cos(\beta z), 
\ee
the mean correction  can now be expressed  as 
\begin{align}
x_{r2}^{2D} = 
\left<\Ut_c | \mathbf{\widetilde{S}}_{2,\Ut_c} \Ut_c \right> 
&=  \left< \tilde{y}_1  \partial_y {U}_0(y_0) |  \mathbf{\widetilde{S}}_{2,\Ut_c} \partial_y {U}_0(y_0)  \tilde{y}_1  \right> \nonumber \\
&=  \left< \tilde{y}_1 |\mathbf{M}^{\dagger} \mathbf{\widetilde{S}}_{2,\Ut_c} \mathbf{M} \tilde{y}_1  \right> = \left< \tilde{y}_1 | \mathbf{\widetilde{S}}_{2,{\tilde{y}}_1}  \tilde{y}_1  \right>,
\end{align}
where $\BF{M} $ is a weight matrix accounting for the wall shear stress $\partial_y U_0(y_0)$ of the uncontrolled flow.
Finally, the optimization for wall-normal deformation reads
\begin{equation}
\min_{||{\tilde{y}_1}||=1} \left( x_{r2}^{2D} \right) =\min \frac{\left< \tilde{y}_1 \left| \frac{1}{2} \left(\widetilde{\mathbf{S}}_{2,\tilde{y}_1}+\widetilde{\mathbf{S}}^{T}_{2,\tilde{y}_1} \right)  \tilde{y}_1  \right.  \right>}{\left<\tilde{y}_1 \left| \tilde{y}_1 \right.  \right>} = \frac{1}{2} \lambda_{\min} \left(\widetilde{\mathbf{S}}_{2,\tilde{y}_1}+\widetilde{\mathbf{S}}^{T}_{2,\tilde{y}_1} \right).
\label{eq:opyc}
\end{equation}

\section{Numerical method} \label{SecNM}
\subsection{Linear analysis and optimization}

The sensitivity analysis and the optimization are conducted using the method described in \citep{Boujo14, Boujo15, Boujo19}.  
The problem is discretized with a finite-element method using FreeFem++ \citep{Hecht12} with P2 and P1 Taylor-Hood elements for velocity and pressure, respectively. 
Mesh points are clustered near the reattachment point, yielding a typical number of elements of $1.6 \times 10^5$ and $10^6$ degrees of freedom. 
The uncontrolled base flow (\ref{eq:OrderE0}) is obtained with a Newton method. 
Eigenvalues are solved with a restarted Arnoldi method. 

At the inlet ($x=-5$), a Poiseuille flow profile is imposed with  maximum velocity $U_{in}=1$, and a stress-free 
condition is applied at the outlet ($x=50$).
At $Re=500$, the reattachment location on the lower wall is $x_{r0}=10.87$ (recall 
$Re = U_{in} h / \nu$  with $h=1$ the step height and $\nu$ the kinematic viscosity).
It is well converged: $x_{r0}=10.88$ on a coarser mesh with $4.5 \times 10^4$  elements. 

\subsection{Three-dimensional DNS}

Direct numerical simulations (DNS) are also carried out for validation of the optimization method, using the open-source code NEK5000 \citep{nek5000}. 
This  parallel code is based on the spectral element method where spatial domain is discretized using hexahedral elements. The unknown parameters are obtained using $N$th-order Lagrange polynomial interpolants, based on the Gauss-Lobatto-Legendre quadrature points in each spectral element with $N\geq6$. A third order backward differentiation formula (BDF3) is employed for time discretization. 
For the spatial discretization, the diffusive terms are treated implicitly whereas the convective terms are estimated using a third order explicit extrapolation formula (EXT3). Since the explicit extrapolations of the convective terms in the BDF3-EXT3 scheme enforce a restriction on the time step for iterative stability \citep{Karniadakis91}, we chose the time step so as to have a Courant number CFL $\approx 0.5$. 

The computational domain and the boundary conditions are in accordance with the specifications of the BFS used in the sensitivity analysis. 
Additionally, we impose periodic boundary conditions in the spanwise direction, where the spanwise width $z \in [0\ 2\pi/\beta]$ captures  one wavelength  for the purpose of validation. Certain cases employing optimal spanwise modulation required the analysis of a  domain with two wavelengths, $z \in [0\ 4\pi/\beta]$. 
The domain is discretized with a structured multiblock grid consisting of 36200 and 72400 spectral elements for the spanwise widths $2\pi/\beta$ and $4\pi/\beta$, respectively.  {In both cases, the minimum and maximum distances between the adjacent grid points are $2.4\cdot 10^{-3}$ (near the step corner and the reattachment point) and $2.2\cdot 10^{-1}$ (at the outlet), respectively.}

\section{Linear stability properties of the 2D uncontrolled base flow}\label{SecGS}

In this section, we  investigate the characteristics of the uncontrolled base flow. 
The BFS flow separates at the step corner and reattaches downstream, thus forming  a recirculation region. 
For the BFS of expansion ratio 2 at $Re=500$, there  {are} two recirculation regions: one on the lower wall developing for $x \in [0\ 10.87]$, and another one on the upper wall for $x \in [8.7\ 17.5]$. 
In this section, we discuss some linear characteristics of the uncontrolled 2D base flow.

\subsection{Global linear stability}

We first investigate the eigenvalues of the system. 
We assume  normal mode perturbations $\qq' = \widehat\qq(x,y) \exp(\lambda t + \mathrm{i} \beta_0 z)$ of small-amplitude,  complex eigenvalue $\lambda$, and real spanwise wavenumber $\beta_0$. 
We use the subscript $_0$ to denote the eigenmode wavenumber  (to be distinguished from the control wavenumber $\beta$).
We solve the generalized eigenvalue problem 
\begin{equation}
   \lambda \widehat{\BF{q}} = {\widetilde \AAA_0} \widehat{\BF{q}}
\end{equation} 
associated with the linearized equation for perturbations around the uncontrolled 2D base flow, with no-slip boundary conditions at the walls. 

\begin{figure}
\centerline{\includegraphics[width=0.6\textwidth]{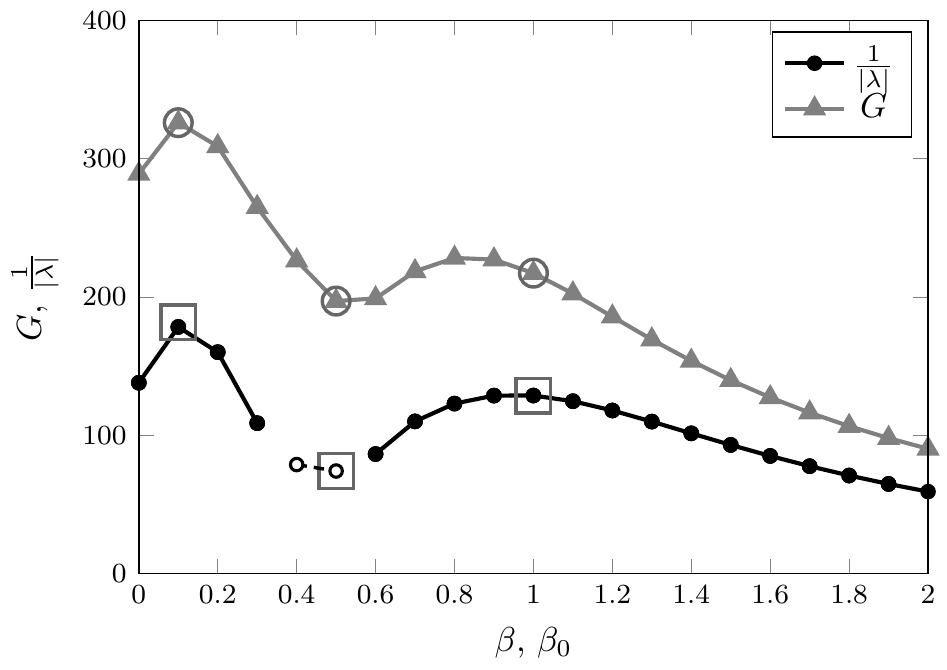}}
\caption{Leading eigenvalue (inverse distance from the origin $1/|\lambda|$) and steady optimal gain $G$, as a function of spanwise wavenumber. 
Filled circles: steady modes (zero frequency $\lambda_i=0$); 
empty circles: oscillating modes (non-zero frequency). 
Highlighted wavenumbers: see Figs.~\ref{fig:eigvec}-\ref{fig:optivec}.
}
\label{fig:growth_gain3Dop}
\end{figure}

Leading eigenvalues for $Re=500$ are shown in Fig.~\ref{fig:growth_gain3Dop} as a function of the spanwise wavenumber $\beta_0$. 
For the purpose of later comparison, we plot the inverse of the absolute value of $\lambda$. 
For all wavenumbers,  the leading eigenvalue has a negative growth rate (stable, decaying modes), and zero frequency (steady modes; filled circles) except near $\beta_0 = 0.4-0.5$ (oscillating modes; empty circles). 
There are two local maxima of $1/|\lambda|$ (least stable modes) near $\beta_0=0.1$ and $\beta_0=1$,
in line with the results of \cite{Barkley02} for $Re=450$. 
%

\begin{figure}
\centerline{\includegraphics[width=\textwidth]{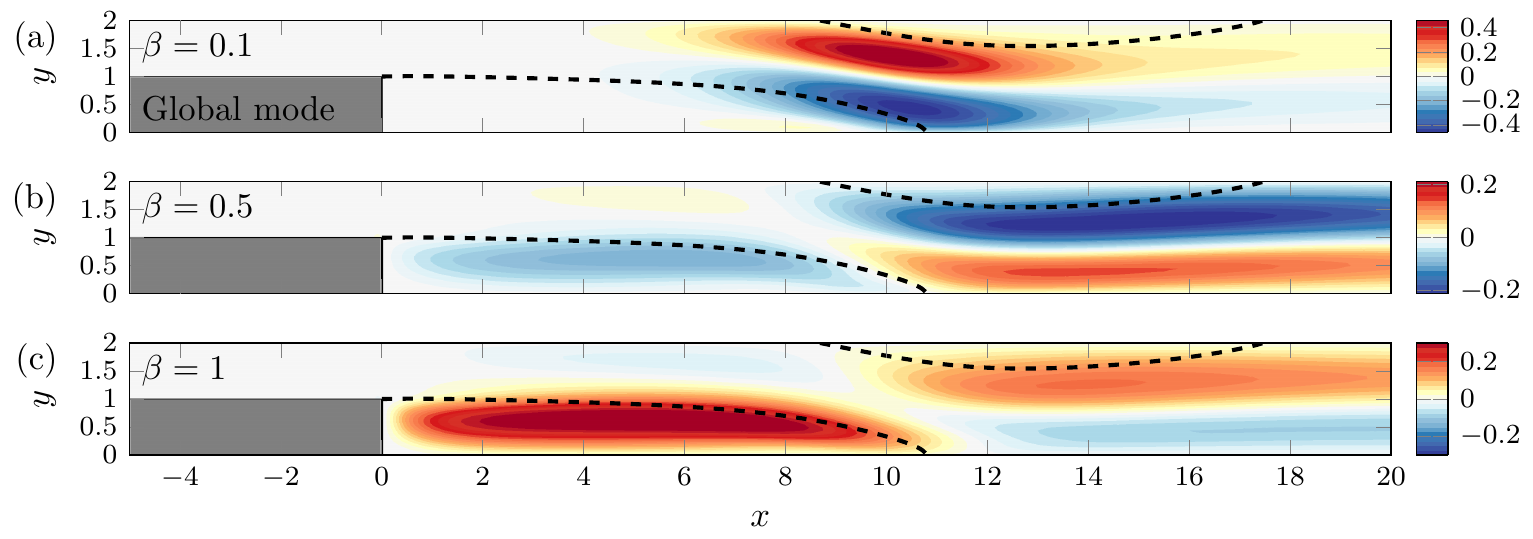}}
\caption{
Streamwise velocity of the 
least stable global eigenmode for 
(a)~$\beta_0=0.1$, 
(b)~$\beta_0=0.5$ and 
(c)~$\beta_0=1$. 
In (a) and (c) $\hat u$ is represented (steady modes) while in (b) the real part  $\Real({\hat u})$  is shown (oscillating mode).
}
\label{fig:eigvec}
\end{figure}

Some selected global modes are shown in Fig.~\ref{fig:eigvec} for $\beta_0=0.1$, $0.5$ and $1$. 
For $\beta_0=0.1$, the mode is localized around $x=10$,  near the lower reattachment and upper separation points. 
For $\beta_0=0.5$, the mode is largest farther downstream ($x>10$), while for $\beta_0=1$ it is localized in the lower recirculation region $x<10$ . 

\subsection{ Optimal 3D steady forcing }

For linearly stable flows, it is interesting to investigate what kind of  disturbances undergo the largest amplification.
Here we consider in particular a steady spanwise-harmonic forcing $\mathbf{f}=\widehat{\BF{f}}(x,y)\exp(\mathrm{i}\beta z)$ acting on the wall boundaries, and resulting linearly in a steady spanwise-periodic response 
$\mathbf{q}=\widehat{\BF{q}}(x,y)\exp(\mathrm{i}\beta z)$
via
\begin{equation}
   {\widetilde \AAA_0} \widehat{\BF{q}} = \BF{B}_f \widehat{\BF{f}},
\end{equation} 
where $\BF{B}_f$ limits active forcing regions to the walls. 
The linear amplification efficiency can be measured with a linear gain, for instance as the ratio of the norms of the forcing velocity and response velocity:
\begin{equation}
   G = \frac{||\widehat{\BF{q}} ||}{|| \widehat{\BF{f}} ||}.
\label{eq:gainop}
\end{equation}
This ratio can be maximized: the linear optimal gain is given by the largest singular value of the resolvent operator (here with zero frequency) and the optimal forcing  is the associated singular vector  \cite{Garnaud13,Boujo15}. 

\begin{figure}
\centerline{\includegraphics[width=\textwidth]{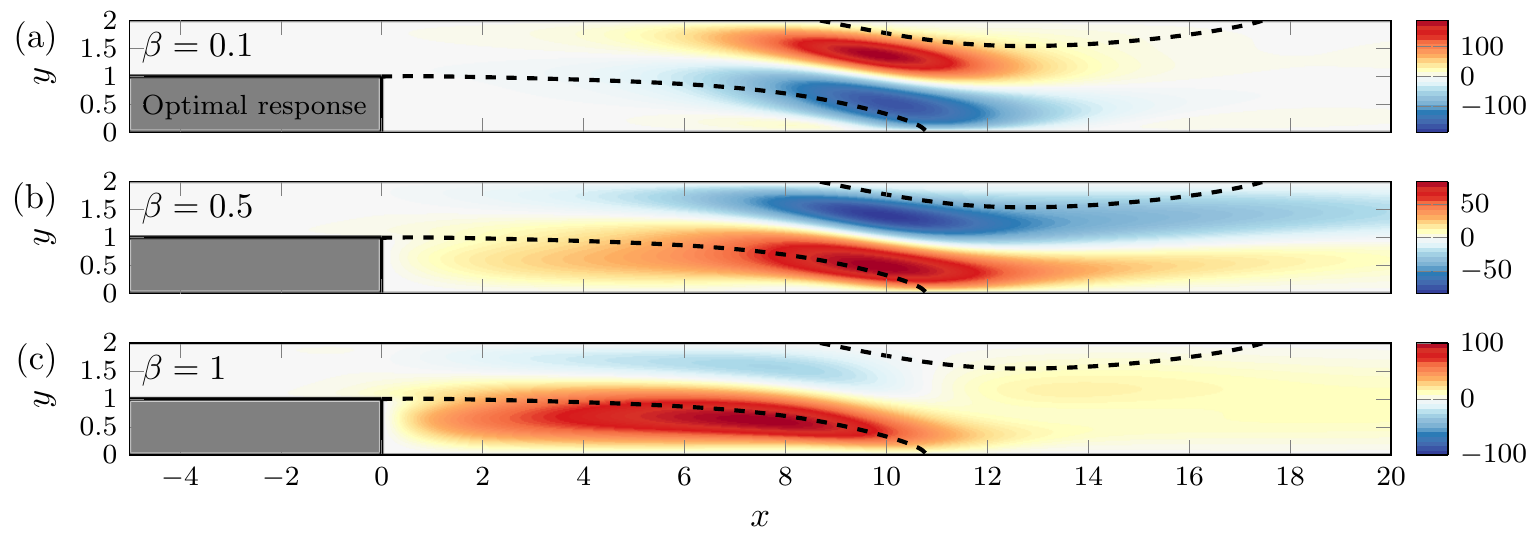}}
\caption{
Streamwise velocity (real part $\Real({\hat u})$) of the
optimal response to steady forcing for 
(a)~$\beta=0.1$,
(b)~$\beta=0.5$ and 
(c)~$\beta=1$.
}
\label{fig:optivec}
\end{figure}

The optimal gain for steady wall actuation is shown in Fig.~\ref{fig:growth_gain3Dop} as function of  the forcing spanwise wavenumber. 
The maximum optimal gain $G=326$ is reached for $\beta=0.1$, the same wavenumber as the least stable eigenmode. 
Qualitatively, the optimal gain varies with the spanwise wavenumber like  $1/|\lambda|$ for the leading global mode. 
This result illustrates the 
$\varepsilon$-pseudospectral property \citep{Trefethen93, Schmid07}.
Some selected optimal responses 
are depicted in Fig.~\ref{fig:optivec}. 
As expected,  the optimal responses for $\beta=0.1$ and $\beta=1$ are similar to the  eigenmodes at the same wavenumbers. 
For $\beta=0.5$, the optimal response is slightly different from the global mode since the latter has a non-zero frequency while the response is steady.

\section{Results: optimal control for lower reattachment location}
\label{SecRe}

We now turn our attention to the optimal spanwise-harmonic control: wall actuation (blowing/suction) in Sec.~\ref{sec:optwallact}, and
wall deformation in Sec.~\ref{sec:optwalldef}.
All results are given for $Re=500$.

\subsection{Optimal wall actuation}
\label{sec:optwallact}

\begin{figure}
\centering
\centerline{\includegraphics[width=\textwidth]{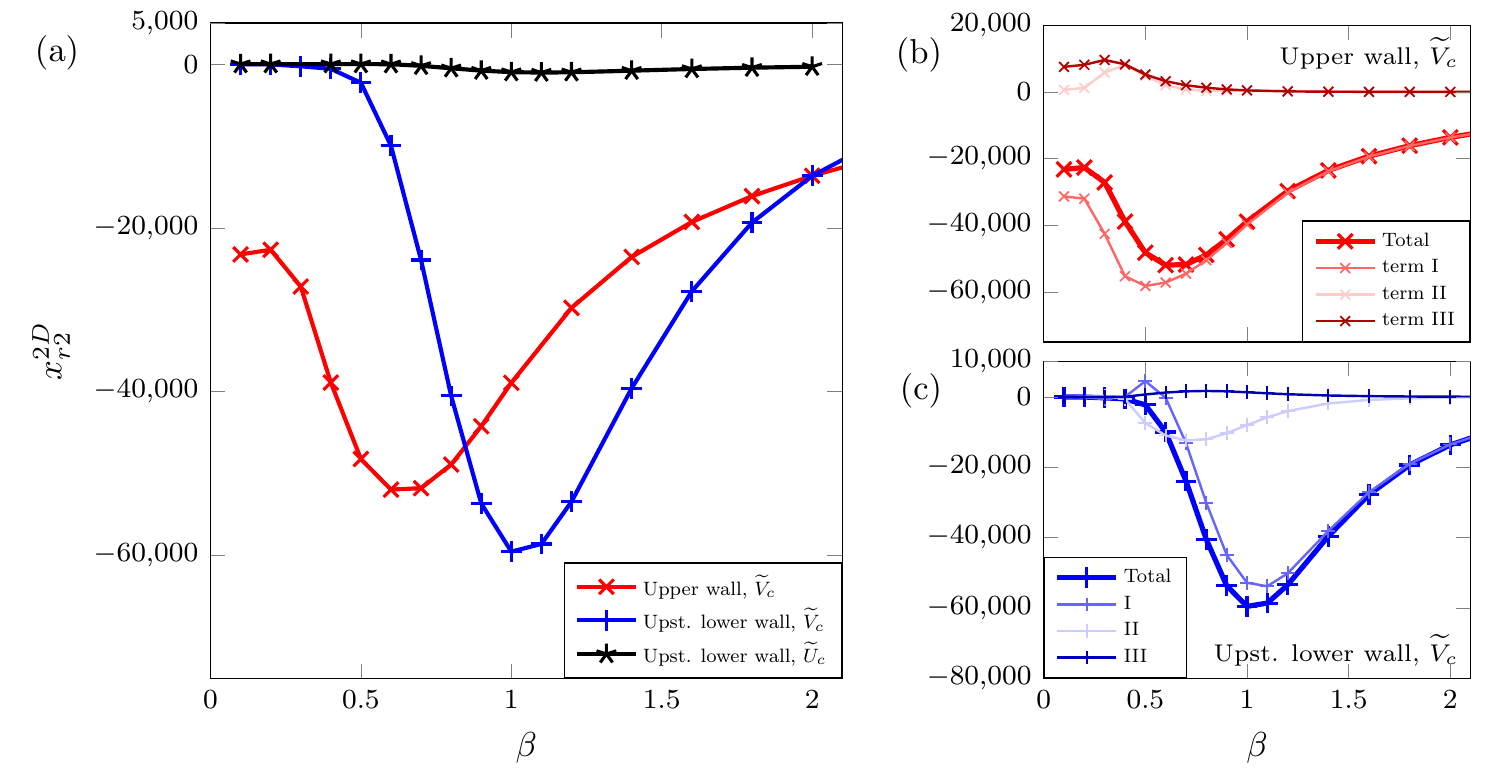}}
\caption{(a)  Mean correction $x_{r2}^{2D}$ induced by the optimal wall blowing/suction minimizing the mean reattachment length $\overline{x_{r}}$ (spanwise wavenumber $\beta$, different walls). The individual contributions of the terms I, II and III in (\ref{eq:xr2_2_simpl}) (their 2D components)  on the total mean correction are detailed in (b) for upper wall, $\Vt_c$ and (c) for upstream lower wall, $\Vt_c$ controls.
}
\label{fig:optimal_xr2d}
\end{figure}

Figure~\ref{fig:optimal_xr2d}(a) shows the optimal negative mean correction $x_{r2}^{2D}$ as a function of $\beta$.
Several wall actuation scenarios are considered: \begin{itemize}
    \item 
    on the upper wall, with  normal velocity $\Vt_c$;
    
    \item 
    on the upstream lower wall, with normal velocity $\Vt_c$;
    
    \item 
    on the upstream lower wall, with tangential velocity $\Ut_c$. 
\end{itemize}
Recall that 3D velocity controls are defined as $(U_c,V_c,W_c)(x,y,z) =(\Ut_c(x,y) \cos(\beta z),$
 $  \Vt_c(x,y) \cos(\beta z), \Wt_c(x,y) \sin(\beta z))$. The wall restriction is implemented by modifying the prolongation matrix $\BF{P}$.

Wall-normal control $\Vt_c$ is most efficient on the upper wall at $\beta=0.6$, and on the upstream lower wall  at $\beta=1$. 
Wall-tangential actuation $\Ut_c$ on the upstream lower wall has a much smaller effect on the reattachment length than normal actuation.
This holds for other types of wall controls (not shown): actuating with  normal velocity $\Vt_c$ is generally more efficient than  with  wall-tangential velocity components  $\Ut_c$ and $\Wt_c$.

\begin{figure}
\centering
\centerline{\includegraphics[width=\textwidth]{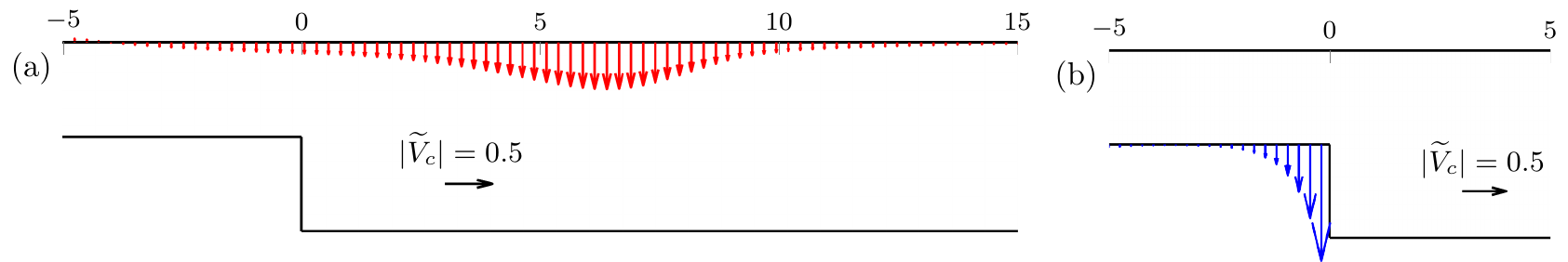}}
\caption{
Optimal control $(0,\Vt_c,0)$ (a) on the upper wall for $\beta=0.6$ and (b) on the upstream lower wall for $\beta=1$. 
}
\label{fig:Op_xr2d_v_b1}
\end{figure}

The individual contributions of terms I, II and III in (\ref{eq:xr2_2_simpl}) 
are shown in Fig.~\ref{fig:optimal_xr2d}(b)-(c) for normal actuation $\Vt_c$ on the upper wall and upstream lower wall, respectively. 
In both cases,  
term I (a linear function of the second-order flow modification) contributes the most on the mean correction, while terms II and III (quadratic functions to the first-order flow modification) have negligible or counteracting  effects. 
Control vectors for the upper wall ($\beta=0.6$) and upstream lower wall ($\beta=1$) are shown in Fig.~\ref{fig:Op_xr2d_v_b1}. 
The control is largest near $x=6$ and $x=0$, respectively.

\begin{figure}
\centering
\includegraphics[width=\textwidth]{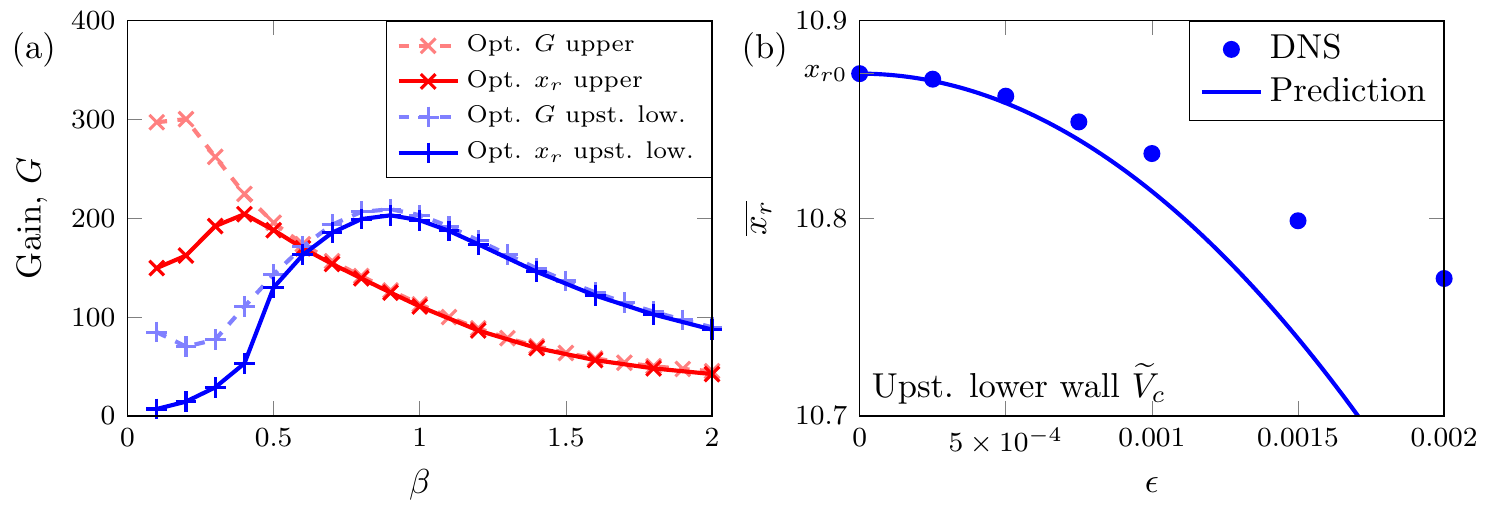}
\caption{(a) Linear gain $G$ for steady  spanwise-periodic wall blowing/suction: 
control $\Vt_c$ minimizing $x_r$ (solid lines)
and 
control $\widehat{\BF{f}}$ maximizing $G$ (dashed lines).
(b)~Mean reattachment location  $\overline{x_{r}}$ as a function of the control amplitude for upstream lower wall actuation for $\beta=1$. 
Line: sensitivity prediction; symbols: 3D DNS. 
}
\label{fig:gain_xr2d}
\end{figure}

The linear gain $G$ for these controls is shown in Fig.~\ref{fig:gain_xr2d}(a) (solid lines).   
Here the gain is calculated as the ratio between the response $||\widetilde \UU_1||$ and the control $||\widetilde \UU_c||$. 
The optimal gain obtained when maximizing  (\ref{eq:gainop}) with wall restriction is also shown in Fig.~\ref{fig:gain_xr2d}(a) (dashed lines).  
The gain obtained by maximizing $x_{r2}$ and $G$ itself are close each other, except for lowest $\beta$ values.
The corresponding flow modifications $\widetilde{\mathbf{U}}_1$ and 
$\widehat{\mathbf{u}}$  (not shown) are very similar to each other too.
This indicates that  the amplification potential of the system is closely related to the recirculation length $x_r$, as reported in \cite{Boujo14b}. 

Figure \ref{fig:gain_xr2d}(b) shows the spanwise-averaged reattachment location $\overline{x}_r$ computed from 3D DNS along with the sensitivity prediction for the reattachment location 
$\overline{x}_r = x_0+\epsilon^2 x_{r2}^{2D}$ 
as a function of the actuation amplitude $\epsilon$, for the upstream lower wall  case. 
The agreement is  good up to $\epsilon \simeq 0.001$. For this amplitude (equal to 0.1\% of the maximum inlet velocity), the optimal control on the upstream lower wall reduces the reattachment location by 0.55\%.
For larger amplitudes in the investigated range, DNS results start to differ due to strong nonlinear effects, but $\overline x_r$  continues to decrease.

\subsection{Optimal wall deformation}
\label{sec:optwalldef}

We now investigate the optimal wall deformation for minimizing the lower reattachment point. 
We focus on the upstream lower wall. 
The wall deformation is computed using (\ref{eq:opyc}),
and we  apply to $y_1$ the smoothing filter
$F_w = 1/(\exp(2 C_k (x+x_S))+1)$, with $C_k=250$ and $x_S=0.02$,
to avoid singularity at the step corner where $\partial_y U_0$ goes to infinity. {This amounts to regularizing the sensitivity (we note that one could also regularize the geometry with a small chamfer at the corner).} 

\begin{figure}
\centering
\centerline{\includegraphics[width=\textwidth]{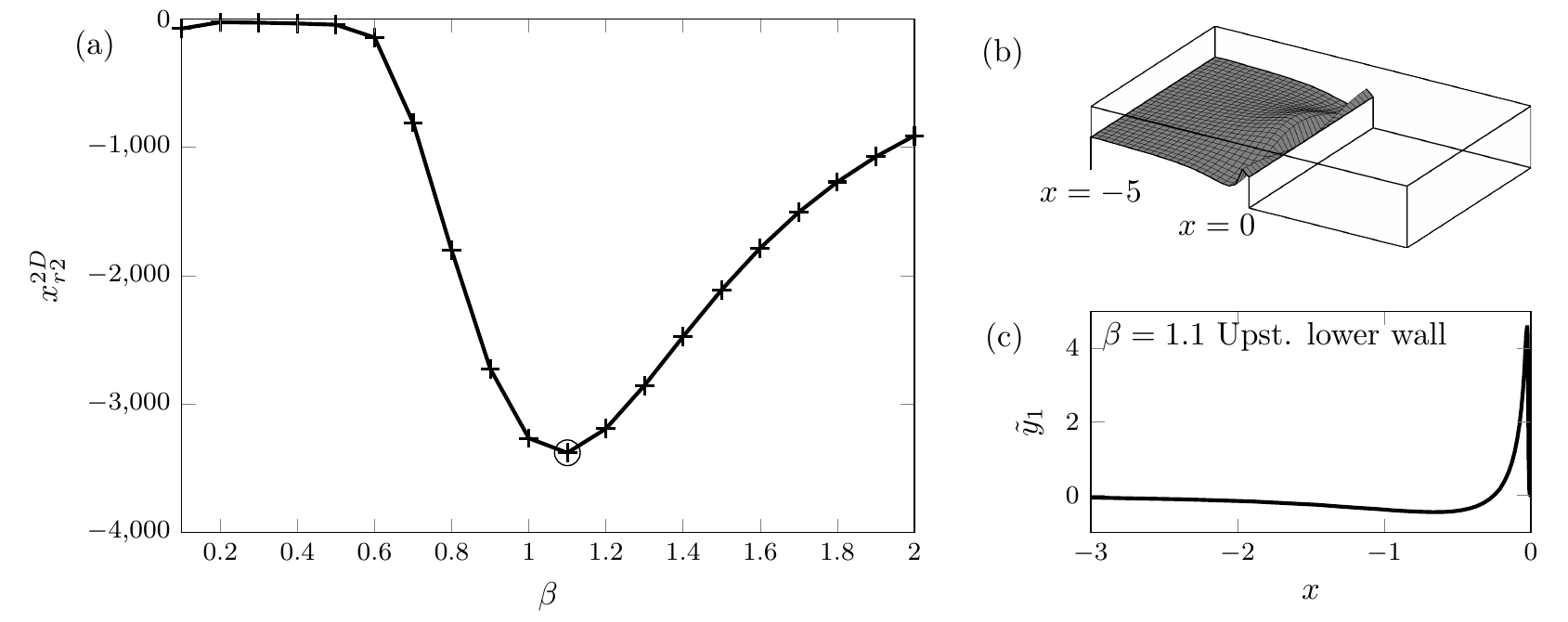}}
\caption{(a) Effect of the optimal upstream lower wall deformation as a function of spanwise wavenumber $\beta$. 
(b)~3D visualization of the optimal upstream lower wall deformation $y_1=\tilde{y}_1 \cos(\beta z)$ and (c)~2D profile $\tilde{y}_1$ for $\beta=1.1$.  
}
\label{fig:optimal_xr2d_Wall}
\end{figure}

Figure~\ref{fig:optimal_xr2d_Wall} shows the effect of the optimal control $x_{r2}^{2D}$ as a function of $\beta$. 
The most effective spanwise wavenumber is $\beta=1.1$, similar to the wall blowing/suction case, 
but the efficiency is much lower (minimum $x_{r2}^{2D}$  about 15 times smaller). 
This is due to the fact that wall deformation is equivalent to a tangential velocity $\Ut_c$, 
which  has a much smaller effect than normal velocity $\Vt_c$ on $x_{r2}$  (recall  Fig.~\ref{fig:optimal_xr2d}). 
Although less effective,  wall deformation on the upstream lower wall still results in the mean correction  $x_{r2}^{2D} = -3.7 \times 10^3$.

\begin{figure}
\centering
\centerline{\includegraphics[width=\textwidth]{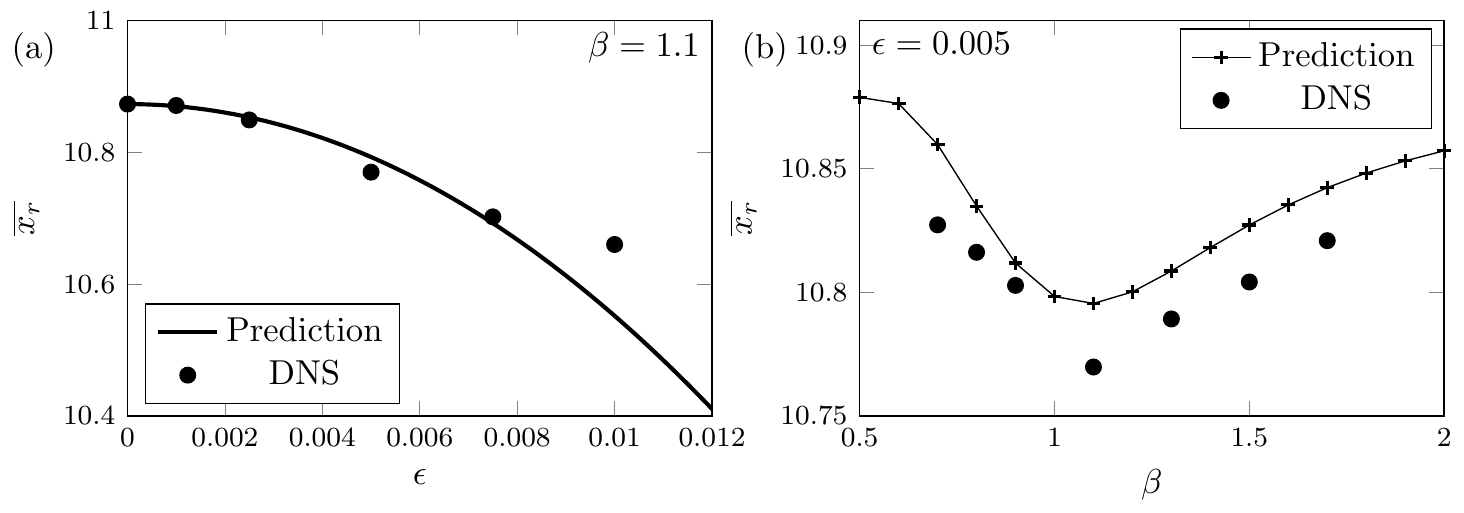}}
\caption{Effect of the optimal wall deformation on the mean reattachment point (a) as a function of $\epsilon$ for fixed $\beta=1.1$ and (b) as a function of $\beta$ for fixed $\epsilon= 0.005$.
}
\label{fig:Optimal_wall_def}
\end{figure}

Figure \ref{fig:optimal_xr2d_Wall}(b)-(c) show the optimal wall deformation $y_1$ and its 2D profile  $\tilde{y}_1$ (recall $y_1 = \tilde{y}_1 \cos (\beta z)$). 
The wall deformation is  maximum just before the step corner, where the flow separates. 
The mean reattachment location from 3D DNS is shown in Fig.~\ref{fig:Optimal_wall_def}(a). 
A good agreement is found until $\epsilon = 0.0075$. 
At this point, $\overline x_r$ is decreased to $10.7$: a deformation amplitude equal to 0.75\% of the inlet channel and step heights reduces the mean reattachment location by 1.5\% . 
For larger deformation amplitudes ($\epsilon>0.01$), DNS results depart from the sensitivity prediction. 

Figure~\ref{fig:Optimal_wall_def}(b) shows $\overline{x}_r$ as a function of $\beta$ for a fixed deformation amplitude $\epsilon=0.005$. 
Overall, sensitivity predictions and 3D DNS results are in good agreement, with a maximum error  $|\overline{x_{r}}_{DNS}-{x}_r^{2D}|/\overline{x_{r}}_{DNS}\simeq 0.2\%$ for $\beta=1.1$.

\begin{figure}
\centering
\centerline{\includegraphics[width=\textwidth]{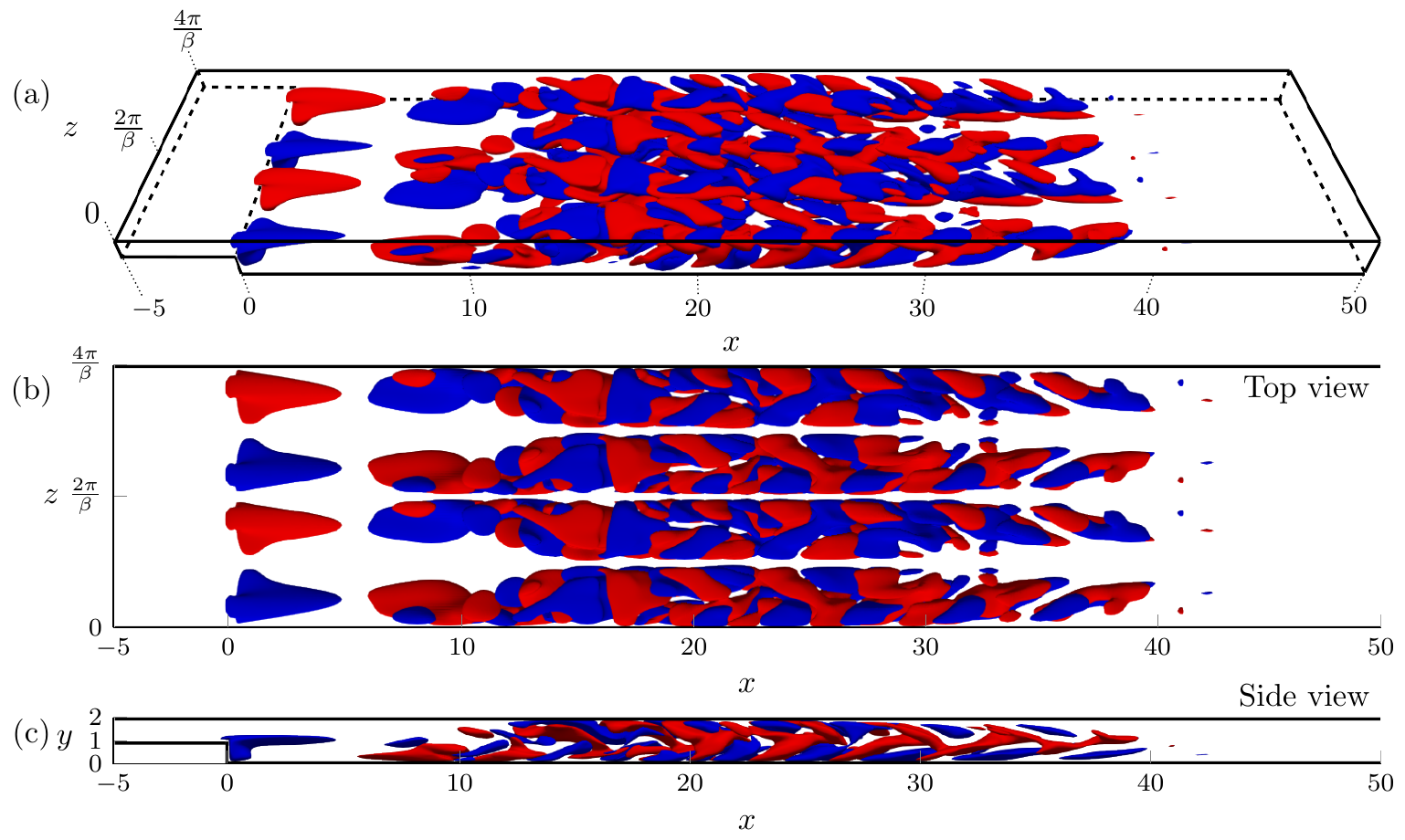}}
\caption{
Iso-surfaces of instantaneous spanwise velocity  $W = \pm 0.03$ for the optimal wall deformation on the upstream lower,  with amplitude $\epsilon=0.015$. 
(a)~oblique view, (b)~top view and (c)~side view.  
}
\label{fig:unstable}
\end{figure}

For a larger deformation amplitude $\epsilon=0.015$, the flow becomes unstable. 
Figure~\ref{fig:unstable} shows an instantaneous flow field with iso-contours of spanwise velocity $W=\pm 0.03$. 
Because the uncontrolled base flow has no spanwise velocity component, $W$ is a good indicator of  velocity perturbations. 
Those perturbations develop just after the step corner and are sustained in the region $x \in [5\ 40]$. 
From the top view in Fig.~\ref{fig:unstable}(b), clear lines of vanishing $W$ are observed at the nodal points of $\sin(\beta z)$. 
Chevron patterns appear in the side view in Fig.~\ref{fig:unstable}(c).  
Perturbations oscillate in time at a fundamental frequency $\omega = 0.55$ ($St=0.088$).  
Boujo, Fani and Gallaire \cite{Boujo15b} reported the destabilizing effect of spanwise-periodic control in parallel shear flow. 
They showed that both  fundamental $\beta$ and sub-harmonic $\beta/2$ modes can be excited due to a sub-harmonic resonance mechanism  \citep{Herbert88,Hwang13}. 
In our DNS with a spanwise domain extended to two control wavelengths ($z \in [0\ 4 \pi/\beta]$), and thus able to 
accommodate perturbations of wavenumber as small as $\beta/2$, perturbations do not show any sub-harmonic component. 
Instead, only harmonics of $n \beta$ ($n=1,2,3...$) exist, as observed in Fig.~\ref{fig:unstable}(b).

\section{Discussion}\label{SecDiscussion}

Although the optimization procedure finds the most efficient spanwise-harmonic control, the effect on the mean recirculation length appears relatively small. 
In light of this observation, it is worth comparing the optimal 2D and 3D blowing/suction. One can show that the optimal 2D wall control is equal to the sensitivity to 2D wall control, given by the adjoint stress at the wall $\left(P^\dag \II + \nnu \bnabla \UU^\dag \right) \nn$, where $(\UU^\dag, P^\dag)$ is the adjoint base flow (see Sec.~\ref{sc:sensit_general}) and $\nn$ the outward unit normal vector \citep{Boujo14, Boujo14b, Boujo15}.  
Since the tangential component is generally much smaller than the normal one, we simply consider the sensitivity to 2D normal actuation as the optimal control $(0,V_c)$.

\begin{figure}
\centering
\centerline{
\includegraphics[height=11cm, trim=5mm 17mm 10mm 5mm, clip=true]{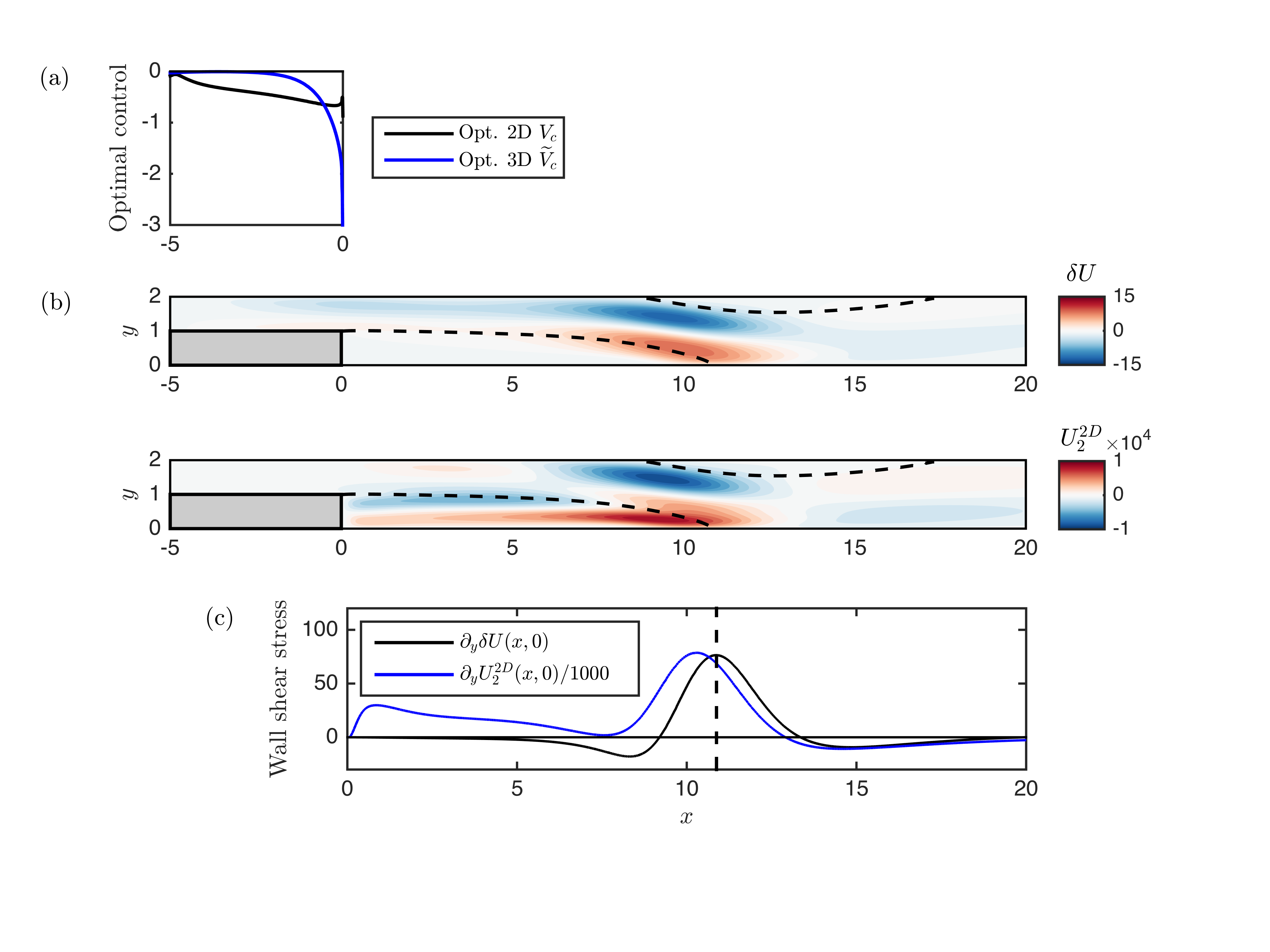}
}
\caption{
(a)~Optimal 2D and 3D ($\beta=1$) vertical controls on the upstream lower wall.
(b)~Leading-order mean flow modifications (streamwise component). 
(c)~Corresponding wall shear stress on the lower wall. 
}
\label{fig:compare_opt_2d_3d_ULW}
\end{figure}

\begin{figure}
\centering
\centerline{
\includegraphics[height=11cm, trim=5mm 17mm 10mm 5mm, clip=true]{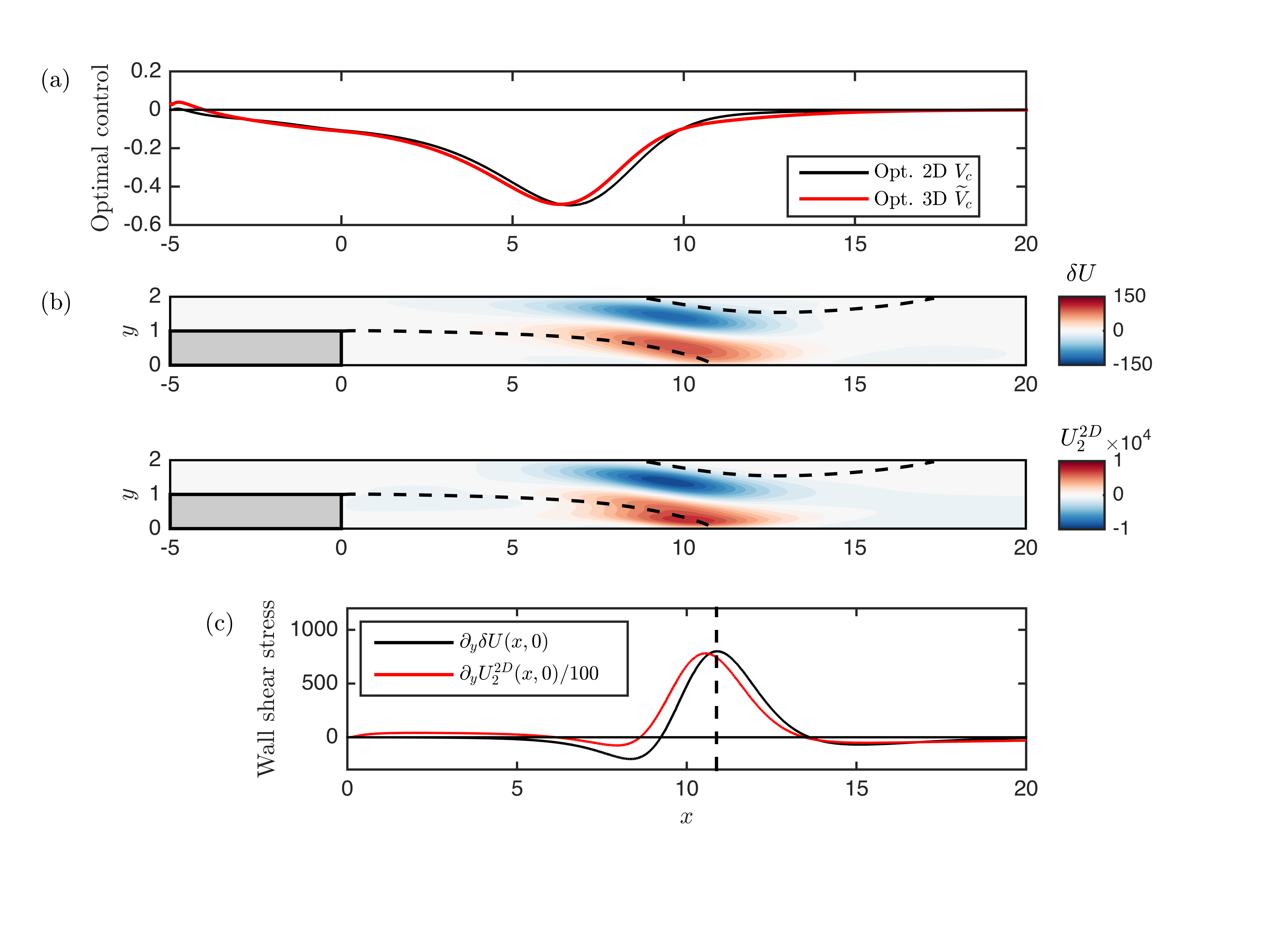}
}
\caption{
(a)~Optimal 2D and 3D ($\beta=0.6$)  vertical controls on the upper wall.
(b)~Leading-order mean flow modifications (streamwise component). 
(c)~Corresponding wall shear stress on the lower wall. 
}
\label{fig:compare_opt_2d_3d_UW}
\end{figure}

\begin{figure}
\centering
\centerline{
\includegraphics[width=7cm, trim=10mm 60mm 20mm 60mm, clip=true]{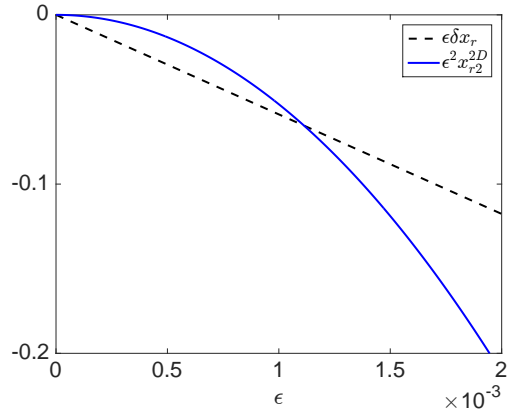}
}
\caption{
Effect on the reattachment location $x_r$ of the optimal vertical 2D control and optimal vertical 3D control ($\beta=1$) of amplitude $\epsilon$, on the upstream lower wall.
}
\label{fig:compare_effect_opt_2d_3d_ULW}
\end{figure}

Figure~\ref{fig:compare_opt_2d_3d_ULW} compares the 3D control optimized on the upstream lower wall ($\beta=1$) to its 2D counterpart, both normalized to  1. 
The linear response $\bdelta \UU$ to the 2D control is largest and positive near the lower reattachment point, resulting in a positive wall shear stress $\partial_y \delta U$ at that location, as expected if $x_r$ is to be minimized. 
 Via the spanwise-periodic first-order flow modification $\UU_1$ (not shown), the optimal 3D control induces a mean second-order flow modification $\UU_2^{2D}$ that is qualitatively similar to $\bdelta \UU$,  resulting  in a positive wall shear stress  $\partial_y U_2^{2D}$, and therefore a negative $x_{r2,\mathrm{I}}$ (we do not investigate $x_{r2,\mathrm{II}}$ and $x_{r2,\mathrm{III}}$ since they are much smaller, as shown in Fig.~\ref{fig:optimal_xr2d}).
Fig.~\ref{fig:compare_opt_2d_3d_UW} shows the same quantities optimized on the upper wall ($\beta=0.6$ for the 3D control), and again a qualitatively similar wall shear stress.
Although $\UU_2^{2D}$ is much larger than $\bdelta \UU$, it must be kept in mind that 2D and 3D controls of the same amplitude $\epsilon$ yield a 2D modification that scales 
linearly ($\sim \epsilon \bdelta \UU$) and a 3D modification that scales 
quadratically ($\sim \epsilon^2 \UU_2^{2D}$), respectively. 
Spanwise-periodic controls should therefore become more efficient for large enough amplitudes, as previously observed for flow stabilization \citep{, del2014optimalB, del2014optimalA, del2014stabilizing, Boujo15b}, and as shown in Fig.~\ref{fig:compare_effect_opt_2d_3d_ULW}. 
In practice, when the control amplitude increases, it may happen that the actual efficiency is limited by deviation from the sensitivity prediction (Sec.~\ref{sec:optwallact})
or by the flow becoming linearly unstable (Sec.~\ref{sec:optwalldef}). 
This can be tested on a case-by-case basis, once promising control candidates have been identified. 
In this respect, the concept of second-order sensitivity and the  associated optimization method allow for a systematic exploration of the best candidates for  spanwise-periodic control.

\begin{figure}
\centering
\centerline{\includegraphics[width=\textwidth]{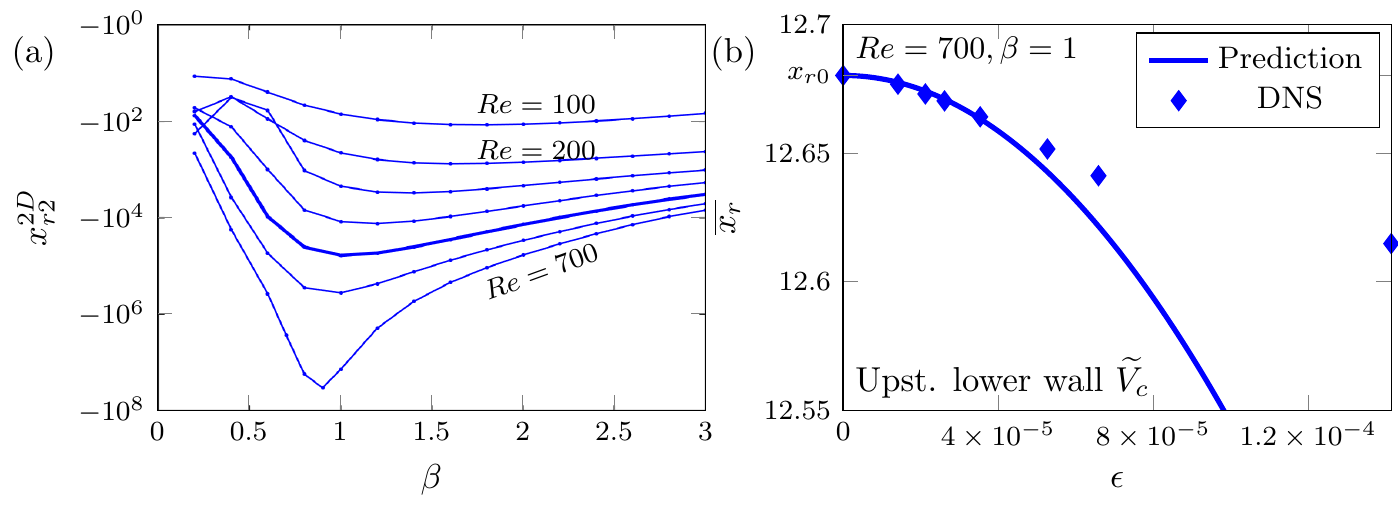}}
\caption{{(a)  Mean correction  $x_{r2}^{2D}$  induced by the optimal wall blowing/suction $\Vt_c$ minimizing the mean reattachment length $\overline{x_{r}}$ (spanwise wavenumber $\beta$, upstream lower wall) for  $Re=100, 200, ... ,700$ with the interval $\Delta Re=100$. The thick line indicates $Re=500$. (b) Mean reattachment location  $\overline{x_{r}}$ as a function of the control amplitude for upstream lower wall actuation for $\beta=1$ and $Re=700$. 
Line: sensitivity prediction; symbols: 3D DNS.} }
\label{fig:ss}
\end{figure}

{This study has focused on $Re=500$. In order to investigate the effect of the Reynolds number, the optimal control has also been computed for other Reynolds numbers up to $Re=700$ (just below the 3D instability threshold).
Figure {\ref{fig:ss}}(a) shows the second-order variation $x_{r2}^{2D}$ for the optimal vertical blowing/suction $\Vt_c$ on the upstream lower wall. The mean correction reaches a maximum for a peak wavenumber that slightly decreases with $Re$, but remains close to $\beta=1-1.5$. The largest mean correction increases exponentially with $Re$. For instance at $\beta=1$, the mean correction for $Re=700$ ($x_{r0}=12.68$) is $x_{r2}^{2D}=-1.35 \cdot 10^7$, which is between two and three orders of magnitude larger than for $Re=500$ $(x_{r0}=10.88)$: $x_{r2}^{2D}=-5.95 \cdot 10^4$. This exponential increase in control authority is similar to the exponential increase in optimal transient growth \mbox{\citep{Blackburn08}} and optimal harmonic gain \mbox{\citep{Boujo15}}, and can be ascribed to the exponential increase in amplification via a shear mechanism, itself related to the linear increase in recirculation length (e.g. \mbox{\citep{Barkley02}}). We note that the profile of the optimal control is very similar at $Re=500$ (Fig. {\ref{fig:Op_xr2d_v_b1}}b) and $700$ (not shown).
Figure {\ref{fig:ss}}(b) shows a DNS validation for $Re=700$, $\beta=1$. The effect is indeed much stronger than for $Re=500$ (Fig. {\ref{fig:gain_xr2d}}b) but higher-order effects appear at a smaller control amplitude.}

\section{Conclusion}\label{SecCo}

Initially motivated by the link between recirculation length and stability properties 
in separated amplifier flows, we have focused on the mean reattachment location as an indicator for the noise amplifying potential in a 3D backward facing step of expansion ratio of 2 and fixed Reynolds number $Re=500$.
In this context, our goal was to control the reattachment location on the BFS lower wall with optimal spanwise-periodic control (steady wall blowing/suction or wall deformation) based on the second-order sensitivity analysis introduced by \cite{Boujo19}  for the linear stability properties of the circular cylinder flow.

A second-order sensitivity tensor for the reattachment location has been derived, such that  modification of the reattachment location is obtained as a scalar product of this  tensor and any arbitrary control.
For the specific case of spanwise-harmonic control, the sensitivity tensor was then further simplified, i.e. made independent of $z$.    
When the control is spanwise harmonic, the first-order reattachment modification takes the same wavenumber with zero mean value, while the second-order modification has a non-zero mean value. Thereby, we have looked for optimal controls that minimize the second-order mean correction. 

For wall  blowing/suction, we have shown that   tangential control has a negligible influence while normal control  is the most effective. The optimal wavenumber $\beta$ depends on the control location: $\beta=0.6$ is optimal when controlling on the upper wall, and $\beta=1$ when controlling on the upstream lower wall control. 
The linear gain for this  actuation resembles  the optimal gain for 3D steady forcing, indicating that the amplification potential of the BFS is indeed linked to the recirculation length, as also observed in \cite{Boujo15}.
Three-dimensional direct numerical simulations have  validated the quadratic behaviour of the mean reattachment length modification. 
The sensitivity prediction is valid until a control amplitude  $\epsilon \simeq 0.001$;  
for larger amplitudes, DNS results start to deviate from the quadratic prediction. 

Optimal wall deformation has  been studied too. 
We have focused on deformation of the upstream lower wall,  restricting the wall deformation to be null at the step corner. 
The optimal wall control  is generally less effective than wall  optimal blowing/suction, and its optimal wavenumber is $\beta=1.1$. 
DNS validation has shown that the sensitivity prediction is valid until a deformation amplitude $\epsilon \simeq 0.008$;
beyond that, the optimal control destabilizes the flow. 

Finally, the optimal 3D spanwise-periodic control 
was compared to the optimal 2D control. 
The resulting wall shear stress (directly linked to the modification of the reattachment location) is two or three orders of magnitude larger for 3D controls than for 2D ones. 
Since 2D and 3D controls depend linearly and quadratically on the  control amplitude, respectively, the 3D control is more efficient  for large enough control amplitudes. 
In order to determine which of the two controls is best at which amplitude, additional studies are required
once the optimal 3D control has been identified. 
This limitation can be tackled if 
the mean flow modification is taken into account in the optimization, for instance with a semi-linear approach
\citep{Vlado14,Meliga16}.   

We have not systematically investigated the stability of the controlled flow. 
Although the spanwise-periodic first-order flow modification does not induce any mean variation of $x_r$, it may still alter the flow stability.
Clarifying whether this is the case or not would be possible, for a given control, using  linear stability analysis (Floquet or 3D global), or non-linear DNS.



\section*{Acknowledgments}
The authors are grateful to Dr. Lorenzo Siconolfi for his help with the direct numerical simulations. 

\appendix
\section{Appendix: Second-order reattachment location modification}
\label{sec:app_xr2}

Recall the definition of the reattachment location \citep{Boujo14, Boujo14b, Boujo15}:
\begin{align} 
x_r &= 
\int_0^\infty H \left(  -\partial_y U(x,0) \right) \,\mathrm{d}x,
\label{eq:def}
\end{align}
where $H$ is the Heaviside function such that $H(\theta<0)=0$ and $H(\theta>0)=1$.
This expression yields indeed the reattachment location since the wall shear stress $\partial_y U(x,0)$ is negative in the recirculation region.
Hereafter, we omit $y=0$ for brevity.
Substituting
\be
U = U_0 + \epsilon  U_1  + \epsilon^2  U_2 + \Oethree
\ee
into (\ref{eq:def}), one obtains:
\begin{align} 
x_r &= 
\int_0^\infty H \left[  -\partial_y U_0 - \epsilon  \partial_y U_1  - \epsilon^2 \partial_y U_2 + \Oethree \right] \,\mathrm{d}x
\nonumber \\
&= 
\int_0^\infty 
\left\{
H \left(  -\partial_y U_0  \right)
- \left[ \epsilon \partial_y U_1  + \epsilon^2 \partial_y U_2 + \Oethree \right] 
H' \left( -\partial_y U_0 \right)
+ \frac{1}{2} \left[ \epsilon \partial_y U_1 + \Oetwo \right]^2 
H'' \left( -\partial_y U_0 \right)
\right\}
\,\mathrm{d}x
\nonumber \\
&= 
\int_0^\infty 
H \left(  -\partial_y U_0  \right)\,\mathrm{d}x
\nonumber \\ 
& \quad
-\epsilon \int_0^\infty 
\left( \partial_y U_1  \right)
H' \left( -\partial_y U_0 \right)
 \,\mathrm{d}x
\nonumber \\ 
& \quad 
+ \epsilon^2 \int_0^\infty \left\{
\left( -  \partial_y U_2  \right)
H' \left( -\partial_y U_0 \right)
+ \frac{1}{2} \left( \partial_y U_1 \right)^2  
H'' \left( -\partial_y U_0 \right)
\right\}
 \,\mathrm{d}x
 +\Oethree.
\end{align}
The zeroth-order {term} is the reattachment location $x_{r0}$ of the uncontrolled flow.
The first-order term $x_{r1}$ is linear in $U_1$ and is therefore zero when averaging over $z$.
The second-order term contains derivatives of $H$, that can be obtained defining 
$G(x) = H \left(  -\partial_y U(x,0) \right) = H(\theta)$
and using the relations
\begin{align}
G'(x) &= \ddfrac{ \left(  H(\theta) \right)}{x}
=  \ddfrac{H }{\theta} \ddfrac{\theta}{x}
= -H'(\theta) \partial_{xy} U,
\\
G''(x) &= \ddfrac{}{x}\left( -H'(\theta) \partial_{xy} U \right)
\nonumber \\
&= - H'(\theta)  \ddfrac{}{x}\left( \partial_{xy} U \right)
-\ddfrac{\left( H'(\theta) \right)}{x} \partial_{xy} U 
\nonumber \\
&= - H'(\theta) \partial_{xxy} U 
-\ddfrac{^2 H}{\theta} \ddfrac{\theta}{x} \partial_{xy} U 
\nonumber \\
&= - H'(\theta) \partial_{xxy} U
+H''(\theta) \left( \partial_{xy} U \right)^2,
\end{align}
which yields
\begin{align}
H'(\theta) &= -\frac{G'(x)}{\partial_{xy} U}
= \frac{\delta(x-x_r)}{\partial_{xy} U},
\\
H''(\theta)  &= \frac{1}{\left( \partial_{xy} U \right)^2} \left( H'(\theta) \partial_{xxy} U + G''(x) \right)
= \frac{1}{\left( \partial_{xy} U \right)^2} 
\left(  \frac{\delta(x-x_r)}{\partial_{xy} U} \partial_{xxy} U - \delta'(x-x_r) \right),
\end{align}
with $\delta(x)$ the Dirac delta function.
The second-order term thus becomes:
\begin{align} 
x_{r2} &= 
 \int_0^\infty \left\{
\left( -  \partial_y U_2  \right)
H'(\theta_0)
+  
\frac{1}{2} 
\left( \partial_y U_1 \right)^2 H''(\theta_0)
\right\}
\,\mathrm{d}x
\nonumber\\
&= 
 \int_0^\infty \left\{
\left( -  \partial_y U_2  \right)
\frac{\delta(x-x_r)}{\partial_{xy} U_0}
+  
\frac{1}{2} 
\frac{\left( \partial_y U_1 \right)^2 }{\left( \partial_{xy} U_0 \right)^2} 
\left(    \frac{\delta(x-x_r)}{\partial_{xy} U_0} \partial_{xxy} U_0 - \delta'(x-x_r) \right)
\right\}
\,\mathrm{d}x
\nonumber\\
&=
-\frac{\partial_y U_2(x_{r0}) }{\partial_{xy} U_0(x_{r0})}
+ 
\left.
\frac{1}{2} \frac{\left( \partial_y U_1 \right)^2 }{\left( \partial_{xy} U_0 \right)^2}  
\frac{\partial_{xxy} U_0 }{\partial_{xy} U_0} 
\right|_{x_{r0}}
+
\frac{1}{2}
\ddfrac{}{x}  \left[
\frac{\left( \partial_y U_1 \right)^2 }{\left( \partial_{xy} U_0) \right)^2}  \right]_{x_{r0}}
\nonumber\\
&=
\left.
-\frac{ \partial_y U_2 }{ \partial_{xy} U_0 }
\right|_{x_{r0}}
+ 
\left.
\frac{ \left(\partial_y U_1 \right) \left( \partial_{xy} U_1 \right) }{ \left( \partial_{xy} U_0 \right)^2}
\right|_{x_{r0}}
-
\left.
\frac{ \left(\partial_{xxy} U_0\right) \left( \partial_y U_1 \right)^2 }{2 \left( \partial_{xy} U_0 \right)^3}
\right|_{x_{r0}}.
\end{align}

\section{Appendix: Simplification of the sensitivity operators}
\label{sec:app_Simplification}

With a spanwise-periodic control of the form
\be 
\UU_c(x,y,z) =  \left( \begin{array}{c}
\widetilde U_{c}(x,y) \cbz \\ \widetilde V_{c}(x,y) \cbz \\ \widetilde W_{c}(x,y) \sbz
\end{array} \right),
\quad
\CC(x,y,z) =  \left( \begin{array}{c}
\widetilde C_x(x,y) \cbz \\ \widetilde C_y(x,y) \cbz \\ \widetilde C_z(x,y) \sbz 
\end{array} \right),
\label{eq:harm_U_C_A}
\ee
 the 1st-order flow modification is  of the form 
\begin{equation}
\QQ_1(x,y,z) =
\left(\begin{array}{c}
\widetilde U_1(x,y) \cbz \\
\widetilde V_1(x,y) \cbz \\
\widetilde W_1(x,y) \sbz \\
\widetilde P_1(x,y) \cbz
\end{array}\right).
\label{eq:Q1perio}
\end{equation}

Let us consider the first term $x_{r2,\mathrm{I}}$ in 
(\ref{eq:xr2_1})-(\ref{eq:1}).
Given the form of $\QQ_1$, the right-hand side  $-\UU_1 \bcdot \bnabla\UU_1$ of (\ref{eq:OrderE2})
is the sum of 2D and 3D terms:
\begin{align}
\ff^{2D}(x,y) &= -\frac{1}{2} 
\left(\begin{array}{c}
(\widetilde U_1 \ddd_x + \widetilde V_1 \ddd_y - \beta \widetilde W_1 ) \widetilde U_1 
\\ 
(\widetilde U_1 \ddd_x + \widetilde V_1 \ddd_y - \beta \widetilde W_1 ) \widetilde V_1
\\ 
0
\end{array}\right),
\\ 
\ff^{3D}(x,y,z) &= -\frac{1}{2} 
\left(\begin{array}{c}
(\widetilde U_1 \ddd_x + \widetilde V_1 \ddd_y + \beta \widetilde W_1 ) \widetilde U_1 \cos(2\beta z)
\\ 
(\widetilde U_1 \ddd_x + \widetilde V_1 \ddd_y + \beta \widetilde W_1 ) \widetilde V_1  \cos(2\beta z)
\\
(\widetilde U_1 \ddd_x + \widetilde V_1 \ddd_y + \beta \widetilde W_1 ) \widetilde W_1  \sin(2\beta z)
\end{array}\right).
\end{align}
The spanwise-harmonic forcing  $\ff^{3D}(x,y,z)$ induces a 3D spanwise-harmonic response $\QQ_2^{3D}(x,y,z)$ that yields a zero-mean variation $x_{r2,\mathrm{I}}^{3D}(z)$.
By contrast, the 2D forcing term $\ff^{2D}(x,y)$ induces the 2D response
\be 
\QQ_2^{2D}(x,y) = 
\left(\begin{array}{c}
 U_2^{2D}(x,y)  \\
 V_2^{2D}(x,y)  \\
0 \\ 
 P_2^{2D}(x,y)
\end{array}\right)
\ee
that yields a non-zero mean  $x_{r2,\mathrm{I}}^{2D}$.
Recalling (\ref{eq:2}), one can therefore write
\begin{align}
x_{r2,\mathrm{I}}^{2D} &= \ps{ \UUa }{ \ff^{2D} }
\\
&= -\dfrac{1}{2} \iint 
\Ua (\widetilde U_1 \ddd_x + \widetilde V_1 \ddd_y - \beta \widetilde W_1 ) \widetilde U_1 
+ \Va (\widetilde U_1 \ddd_x + \widetilde V_1 \ddd_y - \beta \widetilde W_1 ) \widetilde V_1 
\\
&= -\dfrac{1}{2} \iint 
\widetilde U_1 ( \Ua \ddd_x \widetilde U_1  + \Va \ddd_x \widetilde V_1  - \beta \widetilde W_1 \Ua) 
+ \widetilde V_1 ( \Ua \ddd_y \widetilde U_1  + \Va \ddd_y \widetilde V_1  - \beta \widetilde W_1 \Va) 
\\
&= \ps{\widetilde\UU_1} {\widetilde {\mathbf{S}}_{\mathrm{I}'} \widetilde\UU_1},
\end{align}
where the simplified second-order sensitivity operator
\begin{align}
\widetilde {\mathbf{S}}_{\mathrm{I}'} = -\frac{1}{2}
 \left[\begin{array}{ccc}
 \Ua \partial_x &  \Va \partial_x & 0  \\
 \Ua \partial_y &  \Va \partial_y & 0  \\
-\beta \Ua      & -\beta \Va      & 0  
\end{array}\right]
\label{eq:Ktilde}
\end{align}
can be seen formally as a 2D restriction of the operator $\UUa \bcdot \bnabla()^T$.

Let us now consider the second and third terms $x_{r2,\mathrm{II}}$ and $x_{r2,\mathrm{III}}$  in 
(\ref{eq:xr2_1})-(\ref{eq:1}).
Given (\ref{eq:Q1perio}), it is straightforward to show that 
\begin{align}
x_{r2,\mathrm{II}}^{2D} =
\ps{ \widetilde\UU_1 }{ \widetilde{\mathbf{S}}_\mathrm{II} \widetilde\UU_1},
\qquad
x_{r2,\mathrm{III}}^{2D} =
\ps{ \widetilde\UU_1 }{ \widetilde{\mathbf{S}}_\mathrm{III} \widetilde\UU_1},
\end{align}
where the simplified second-order sensitivity operators are
   %
%
\begin{align}
   \widetilde {\mathbf{S}}_\mathrm{II} &= 
   \dfrac{1}{2 \left(\partial_{xy} U_0(x_{r0})\right)^2 }  
   \delta(x_{r0})
   \left( \ex \partial_y \right)^\dag
   \otimes
   \left( \ex \partial_{xy} \right),
   \label{eq:SII_t}
\\
   \widetilde {\mathbf{S}}_\mathrm{III} &= 
   \frac{ - \partial_{xxy} U_0(x_{r0})  }{4 \left( \partial_{xy} U_0(x_{r0}) \right)^3}
   \delta(x_{r0})
   \left( \ex \partial_y \right)^\dag
   \otimes
   \left( \ex \partial_y \right),
   \label{eq:SIII_t}
\end{align}

Finally, the mean second-order variation is
\begin{align}
x_{r2}^{2D} &= \ps{\widetilde\UU_1} {\widetilde{\mathbf{S}}_{2,\widetilde\UU_1} \widetilde\UU_1}
\quad 
\mbox{where} 
\quad
\widetilde{\mathbf{S}}_{2,\widetilde\UU_1} = \widetilde{\mathbf{S}}_\mathrm{I'} + \widetilde{\mathbf{S}}_\mathrm{II} + \widetilde{\mathbf{S}}_\mathrm{III},
\end{align}
and
 the second-order sensitivities to  control defined by (\ref{eq:S2_C_Uc}) 
 read
\begin{align}
\widetilde{\mathbf{S}}_{2,\widetilde\CC} 
&=\PP^T   
\left.{\widetilde{\AAA}_{0,\widetilde \CC} }^\dag \right.^{-1}
\widetilde{\mathbf{S}}_{2,\widetilde\UU_1}  
\left.{\widetilde{\AAA}_{0,\widetilde \CC} } \right.^{-1}
\PP
\quad (\mbox{volume-forcing-only } {\widetilde{\AAA}_{0,\widetilde \CC} }),
\label{eq:S2Ctilde}
\\
\widetilde{\mathbf{S}}_{2,\widetilde\UU_c} &=  \PP^T 
\left.{\widetilde{\AAA}_{0,\widetilde \UU_c} }^\dag \right.^{-1}
\widetilde{\mathbf{S}}_{2,\widetilde\UU_1} \left.{\widetilde{\AAA}_{0,\widetilde \UU_c} } \right.^{-1}
\PP \quad (\mbox{wall-forcing-only } {\widetilde{\AAA}_{0,\widetilde \UU_c} }),
\label{eq:S2Uctilde}
\end{align}
with 
\begin{align}
{\widetilde \AAA_0} =  \left[\begin{array}{cccc}
 U_0\ddd_x +V_0\ddd_y+\ddd_x U_0 - \widetilde D  & \ddd_y U_0 & 0 &   \ddd_x \\
\ddd_x V_0 & U_0\ddd_x +V_0\ddd_y+\ddd_y V_0 - \widetilde D & 0 &   \ddd_y \\
0 & 0 &  U_0\ddd_x +V_0\ddd_y  - \widetilde D & -\beta \\
\ddd_x & \ddd_y & \beta & 0 
\end{array}\right],
\end{align}
 \begin{align}
\widetilde D =  \nnu (\ddd_{xx}+\ddd_{yy}-\beta^2).
\end{align}

\bibliography{biblio_all}
\end{document}